\documentclass{article}
\usepackage{geometry}
\usepackage{titlesec}
\usepackage{setspace} % For adjusting line spacing
\usepackage{ragged2e}
\usepackage[colorlinks=true, urlcolor=blue, linkcolor=blue, citecolor=blue, bookmarksnumbered=true, bookmarksopen=true]{hyperref}
\usepackage{graphicx}
\usepackage{caption}
\usepackage{hhline}
\usepackage{fancyhdr}
\usepackage{lastpage}
\usepackage{amsmath}
\usepackage{ifthen}
\usepackage{url}
\usepackage{footmisc}
\usepackage{xcolor}
\usepackage{etoolbox}
\usepackage{wrapfig}
\usepackage{cite}
\let\oldbibliography\thebibliography
\renewcommand{\thebibliography}[1]{%
  \oldbibliography{#1}%
  \setlength{\itemsep}{1pt}%
  \setlength{\parskip}{0pt}%
}
\usepackage{authblk}
\usepackage{lineno}
\usepackage{textcase}
\usepackage{tabularx} % For better control over table column widths
\usepackage{svg}
\usepackage{tikz}
\usepackage{subcaption}   % for subfigures
\newcommand{\revisioncolor}{\textcolor{black}}

\definecolor{lightgraytext}{gray}{0.4} % Adjust the shade of grey as needed
\definecolor{graytext}{gray}{0.4}  % 0 = black, 1 = white

\DeclareCaptionLabelFormat{figlabel}{{FIG. #2. }}
\DeclareCaptionLabelFormat{tablelabel}{{TABLE #2. }}
\titleformat{\section}
  {\bfseries\fontsize{10}{12}} % Set font size and style
  {\thesection.} % Section number
  {0.5em} % Horizontal spacing
  {} % Code before title

\titleformat{\subsection}
  {\fontsize{10}{12}\selectfont\bfseries} % Set font size and style
  {\thesubsection.} % Section number
  {0.5em} % Horizontal spacing
  {} % Code before title

\titleformat{\subsubsection}
  {\fontsize{10}{12}\selectfont\itshape} % Set font size and style
  {\thesubsubsection.} % Section number
  {0.5em} % Horizontal spacing
  {} % Code before title

\newcommand{\centeredparheader}[1]{%
    \textcolor{graytext}{%
        \parbox[c]{\textwidth}{%
            \centering #1%
        }%
    }%
}

\fancypagestyle{plain}{% This is for the first page
  \fancyhf{}%
  \fancyhead[C]{\centeredparheader{%
    \fontsize{8}{0}{H. ZHANG et al.} 
  }}%
  \fancyfoot[C]{\textcolor{graytext}{\thepage/\pageref*{LastPage}}} % ← Add this line
}

\newcommand{\evenheader}{\centeredparheader{%
  \fontsize{8}{0}{H. ZHANG et al.} 
   }}
    
\newcommand{\oddheader}{\centeredparheader{%
    \fontsize{8}{0}{H. ZHANG et al.} 
    }}
\makeatletter
\newcommand{\undersim}[1]{\mathrel{\mathpalette\@undersim{#1}}}
\newcommand{\@undersim}[2]{%
  \vcenter{%
    \ialign{%
      ##\cr
      $\m@th#1#2$\cr
      \noalign{\nointerlineskip\kern.2ex}
      $\m@th#1\sim$\cr
      \noalign{\kern-.4ex}
    }%
  }%
}
\newcommand{\gsim}{\undersim{>}}
\newcommand{\lsim}{\undersim{<}}

\title{\Large\bfseries{Flux pumping and bifurcated relaxations of helical core in 3D magnetohydrodynamic modelling of ASDEX Upgrade plasmas}}

\author[1, $\ast$]{H. Zhang}
\author[1]{M. Hoelzl}
\author[2,3]{I. Krebs}
\author[1]{A. Burckhart}
\author[1]{A. Bock}
\author[1]{S. Guenter}
\author[1]{V. Igochine}
\author[1]{K. Lackner}
\author[4,5]{D. Bonfiglio}
\author[1]{E. Fable}
\author[1]{F. Stefanelli}
\author[6,1]{R. Ramasamy}
\author[1]{H. Zohm}
\author[7]{JOREK TEAM}
\author[8]{ASDEX UPGRADE TEAM}

\affil[1]{Max Planck Institute for Plasma Physics, Garching, Germany}
\affil[2]{Eindhoven University of Technology, Eindhoven, The Netherlands}
\affil[3]{Dutch Institute for Fundamental Energy Research, Eindhoven, The Netherlands}
\affil[4]{Consorzio RFX (CNR, ENEA, INFN, Università di Padova, Acciaierie Venete SpA), Padova, Italy}
\affil[5]{Istituto per la Scienza e la Tecnologia dei Plasmi, CNR, Padova, Italy}
\affil[6]{Proxima Fusion GmbH, Munich, Germany}
\affil[7]{See author list of [\href{https:doi.org/10.1088/1741-4326/ad5a21}{M. Hoelzl et al 2024 Nucl. Fusion 64 112016}]}
\affil[8]{See author list of [\href{https:doi.org/10.1088/1741-4326/ad249d}{H. Zohm et al 2024 Nucl. Fusion 64 112001}]}
\affil[$\ast$]{Email: \href{mailto:haowei.zhang@ipp.mpg.de}{haowei.zhang@ipp.mpg.de}}

\begin{document}

% \linenumbers

% \title{\raggedright
% \vspace{-2.2em}
%   \textbf{\fontsize{12}{15}\selectfont
%   % \textcolor{red}{\MakeUppercase{Conference Pre-Print}}\\[2em]
%   % \MakeUppercase{Flux pumping regimes and bifurcated plasma states in MHD modelling of ASDEX Upgrade plasmas}
%   Flux pumping and bifurcated relaxations of helical core in 3D magnetohydrodynamic modelling of ASDEX Upgrade plasmas
%   }
%   \vspace{-3.2em}
%   \setstretch{0.6}
% }

\date{}
\maketitle
% \vspace{1em}

% \begin{flushleft}
%   H. Zhang$^{1,*}$, M. Hoelzl$^1$, I. Krebs$^{2,3}$, A. Burckhart$^1$, A. Bock$^1$, S. Guenter$^1$, V. Igochine$^1$, K. Lackner$^1$, D. Bonfiglio$^{4,5}$, E. Fable$^1$, F. Stefanelli$^1$, R. Ramasamy$^{6,1}$, H. Zohm$^1$, JOREK TEAM$^7$, and ASDEX UPGRADE TEAM$^8$ \\
%   \vspace{0.5em}
%   $^1$Max Planck Institute for Plasma Physics, Garching, Germany \\
%   $^2$Eindhoven University of Technology, Eindhoven, The Netherlands \\
%   $^3$Dutch Institute for Fundamental Energy Research, Eindhoven, The Netherlands \\
%   $^4$Consorzio RFX (CNR, ENEA, INFN, Università di Padova, Acciaierie Venete SpA), Padova, Italy \\
%   $^5$Istituto per la Scienza e la Tecnologia dei Plasmi, CNR, Padova, Italy \\
%   $^6$Proxima Fusion GmbH, Munich, Germany \\
%   $^7$See author list of [\href{https:doi.org/10.1088/1741-4326/ad5a21}{M. Hoelzl et al 2024 Nucl. Fusion 64 112016}]\\
%   $^8$See author list of [\href{https:doi.org/10.1088/1741-4326/ad249d}{H. Zohm et al 2024 Nucl. Fusion 64 112001}]\\
%   $^*$Email: \href{mailto:haowei.zhang@ipp.mpg.de}{haowei.zhang@ipp.mpg.de}
% \end{flushleft}

\section*{\bfseries{Abstract}}
{
% \fontsize{9pt}{12pt}\selectfont\hspace{1cm} 
Flux pumping was achieved in recent hybrid scenario experiments in the ASDEX Upgrade (AUG) tokamak, which is characterized by a sawtooth-free helical quiescent state and the anomalous radial redistribution of toroidal current density and poloidal magnetic flux. \revisioncolor{In this article, the} self-regulation mechanism of the AUG core plasma during flux pumping is investigated at realistic parameters using the JOREK code based on the two-temperature, nonlinear, full magnetohydrodynamic (MHD) model. A key milestone in AUG flux pumping modelling is achieved by quantitatively reproducing the clamped current density and safety factor profiles in the plasma core, demonstrating the effectiveness of the dynamo effect in sustaining the flux pumping state. The dynamo term, \revisioncolor{that is} of particular interest, is primarily generated by the pressure-gradient driven m/n = 1/1 quasi-interchange-like MHD instability. The work systematically extrapolates the parameter regimes of flux pumping from the above AUG base case by scanning dissipation coefficients and plasma beta. The simulation results reveal bifurcated plasma behaviour\revisioncolor{s at different Hartmann numbers}, including distinct states such as flux pumping (helical core with a flat current density), sawteeth (periodic kink-cycling), single crash (without subsequent cycle), and quasi-stationary magnetic island (peaked current density). Transitions from marginal flux pumping state to sawteeth are observed in long-term simulations. The relationships between system dissipation, plasma beta, and different plasma states are carefully analyzed. \revisioncolor{For practical purposes, the potential operational window for flux pumping, as determined by plasma density and temperature, is estimated.} The modelling efforts advance the understanding of flux pumping and facilitate the development of a fast surrogate model for efficient evaluation of flux pumping.
}

\vspace{0.2cm}

\noindent\textbf{Keywords:} flux pumping, sawtooth, dynamo, hybrid scenario, Hartmann number, ASDEX Upgrade

\renewcommand{\footnoterule}{\hrule width \linewidth}

\section{Introduction}
Sawtooth control is essential for long-pulse operations in next-generation tokamaks like ITER and DEMO \cite{Chapman2013NFITER}. To address this need, the self-regulating flux pumping mainly achieved in the hybrid scenario offers a potentially robust solution \cite{Petty2009PRLfluxpumping, Burckhart2023NF, Mao2023PRR, Burckhart2024EPS, Boyes2024NF, Alex2025IAEA, Sam2025IAEA}. Experimentally, flux pumping is identified by the anomalous radial redistribution of central current density (equivalent to poloidal magnetic flux) during the quiescent sawtooth-free phase with m/n = 1/1 instability (m and n are the poloidal and toroidal mode numbers, respectively) \cite{Burckhart2023NF, Mao2023PRR, Burckhart2024EPS, Boyes2024NF, Alex2025IAEA}. The redistribution clamps the on-axis safety factor ($q_0$) near unity and prevents sawtooth onset. Theoretically, flux pumping has been carefully studied through qualitative nonlinear magnetohydrodynamic (MHD) modelling. Such efforts reveal the critical role of the dynamo effect, predict the positive correlation between the efficiency of flux pumping dynamo and the poloidal plasma beta ($\beta_p$) \cite{Jardin2015PRL, Krebs2017PoPFluxPumping}, and assess the role of bootstrap current in flux pumping in the low-density regime \cite{Yu2024NFfp}. Recent JOREK simulations enable a quantitative comparison between MHD modelling and AUG experiments, showing consistent current density deficit and dynamo electric field during flux pumping \cite{Zhang2025NFfluxpumping}, and establishing the foundation for the next modelling stage of JET flux pumping experiments \cite{Burckhart2024EPS, Alex2025IAEA}. 

In retrospect, nonlinear MHD modelling has proven to be crucial for assessing the operating window of flux pumping in tokamaks. In the 1980s, the m/n = 1/1 convection cell driven by resistivity ($\eta$)/temperature gradient was modelled with a reduced MHD model \cite{Denton1987PoF}, where the sawtooth-free solution is more readily obtained at low heat conductivities ($\chi$) and at large ratios of resistive time to Ohmic-heating time. In the 2010s, different regimes of stationary helicity and sustained kink cycles were investigated with the $\eta\chi$MHD model in the XTOR-2F code \cite{Halpern2011PPCF}, showing a threshold in $\beta_p$ for the stationary solution, with a lower $\chi_\perp$ favoring the kink cycling. A subsequent work of XTOR-2F, incorporating the two-fluid effect, studying the diamagnetic threshold for sawtooth cycling \cite{Halpern2011PoPSawtooth} found that the stationary helical state prefers regimes with weak ion-diamagnetic drift and high resistivity. Recently, the problem was further studied by other codes based on visco-resistive full MHD \revisioncolor{models}. The small sawtooth or the steady state with m/n = 1/1 magnetic island was obtained in both high- \cite{Shen2018NF} and low-viscosity \cite{Zhang2020NFfluxpumping} regimes. The plasma beta threshold for flux pumping was confirmed in \cite{Zhang2020NFfluxpumping}, consistent with \cite{Krebs2017PoPFluxPumping, Halpern2011PPCF}. 

In addition to tokamak plasmas, self-organized states --- namely the single‑helicity (SH) and quasi single‑helicity (QSH) states --- have also been a major research focus for reversed\revisioncolor{-}ﬁeld pinches (RFPs). The SH/QSH states are distinct from the more common multiple helicity (MH) state, which is manifested by periodic sawtooth-like relaxations and a wide spectrum of MHD modes due to magnetic reconnections \cite{Cappello2000PRLRFP, Cappello2006PoPRFP}. Through MHD modelling of the RFX-mod device, different plasma regimes of SH/QSH and MH states were identified and compared with experiments. A key finding is that the SH/QSH state predominates under two distinct conditions: (i) low Hartmann numbers (high system dissipation) with ideal boundary conditions \cite{Cappello2000PRLRFP, Cappello2006PoPRFP}; or (ii) high Hartmann numbers (low system dissipation, more consistent with experiments) with a helical boundary condition \cite{Bonfiglio2013PRL, Veranda2019NFRFX}. The relevant dynamo effect responsible for sustaining the SH/QSH state in RFP was analyzed \cite{Cappello2006PoPRFP} and extrapolated to tokamak \cite{Piovesan2017NF}, yielding a similar profile of dynamo electric field in the laminar helical states of both device types. \revisioncolor{It should be noted that the dynamo effect is an essential feature of the RFP configuration also in the MH regime, as was already evident in the early RFP research \cite{Hartog1999PoPdynamo}.}

The outlined research substantially improved the understanding of self-regulating plasma dynamics and current density redistribution mechanism during flux pumping. However, the approximate flux pumping parameter space remains undetermined, particularly in the low-dissipation regime that is \revisioncolor{directly} relevant to experiments. The uncertainty complicates the experimental search for flux pumping on existing devices and complicates the scenario design for future ones. Based on the achieved quantitative modelling of flux pumping for AUG \cite{Zhang2025NFfluxpumping}, the main objective of this work is to investigate the parameter regimes of flux pumping in AUG within the single-fluid framework, with a particular focus on viscosity, resistivity, and plasma beta. 3D nonlinear modelling predicts the bifurcation of core plasma dynamics into different states, including quasi-stationary flux pumping, sawteeth (kink cycling), flux pumping-sawtooth transition, etc. The findings are generally consistent with existing literature, and they offer insights into the core plasma physics at conditions much closer to experiments. Nevertheless, there is still a noticeable difference in the plasma beta threshold for flux pumping between the single-fluid MHD modelling and AUG experiments, highlighting the need for the ongoing extended modelling including the two-fluid effect \revisioncolor{\cite{Halpern2011PoPSawtooth, antlitz2025comparisonmhdgyrokineticsimulations, King2011PoPRFP, Ding2004PRLMST_Halldynamo}} and kinetic energetic particle (EP) physics (e.g., fishbones) \cite{Burckhart2023NF, Guenter1999FB, Kolesnichenko2007popEP_QSI, Bogaarts2022JOREK_Kinetic, Zhang2025PPCF_EP_kink, Zhang2026NF_ST_EP} to more precisely predict the operating window of flux pumping.

The remainder of this paper is organized as follows. Sec. \ref{Sec3DBaseCase} provides a brief introduction to the recent AUG flux pumping experiment and the latest quantitative MHD modelling results of flux pumping for the corresponding AUG discharge with JOREK. Sec. \ref{Secparameterscans} presents the systematic MHD simulation results for extrapolating the parameter regimes of flux pumping in AUG scenarios. These include the magnetic Prandtl number and the Hartmann number (relevant to resistivity and viscosity), as well as the plasma beta. Bifurcated plasma states, including flux pumping, sawteeth, and flux pumping-sawtooth transition, will be analyzed. \revisioncolor{Sec. \ref{SecExpParameter} discusses the relationship between the estimated flux pumping regimes and the corresponding experimental parameters, with particular focus on the plasma density and temperature.} In the end, the findings of the present work and the outlook for the future work are summarized and discussed in Sec. \ref{secSummary}.

\section{AUG flux pumping discharge and 3D MHD modelling}\label{Sec3DBaseCase}
\begin{figure}[htbp]
\centering
  \begin{tikzpicture}
  \node[inner sep=0pt] (figs) at (0,0)
  {\includegraphics[width=0.5\textwidth]{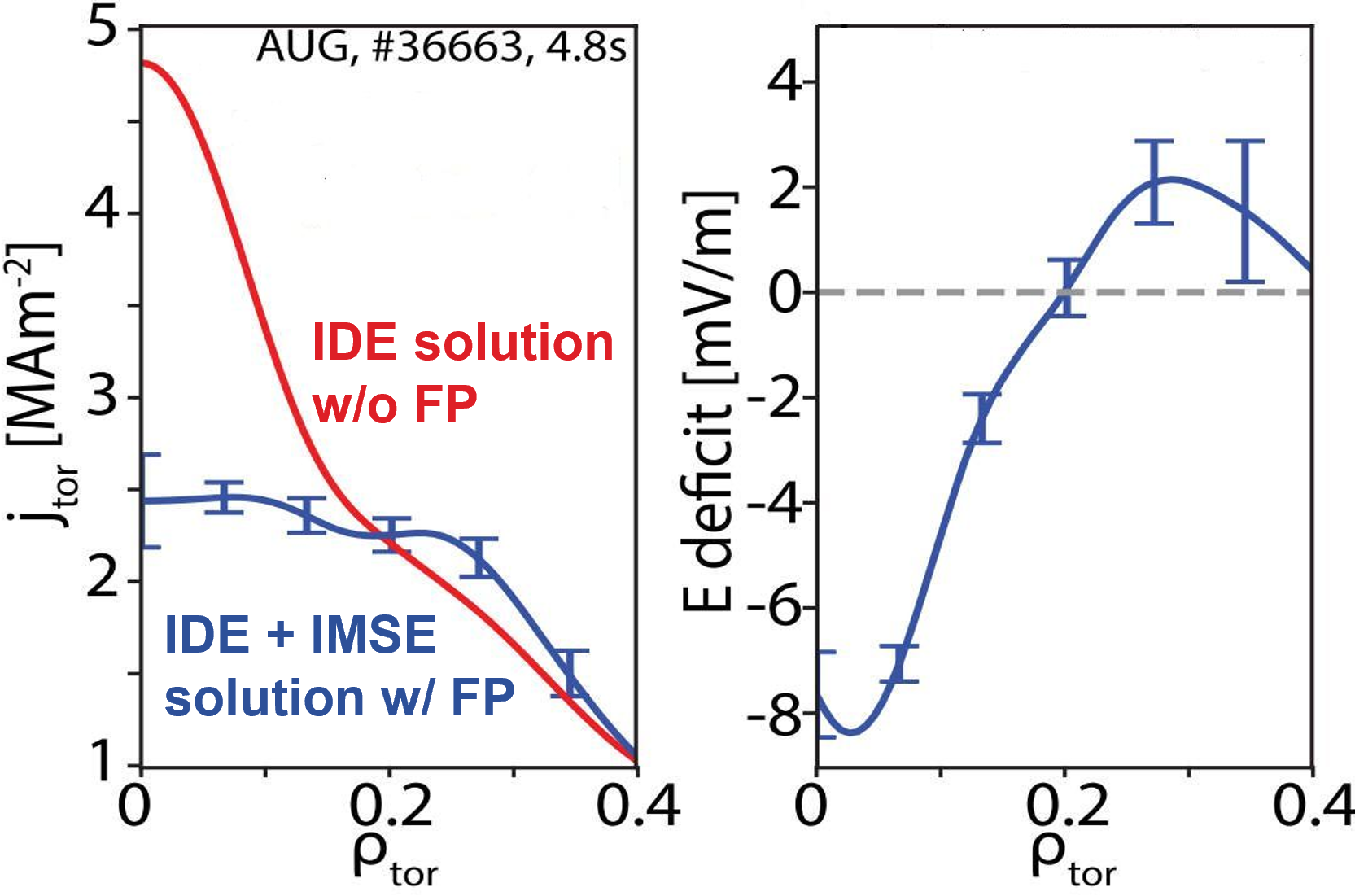}};
  \node[align=center,fill=none] at (-1.8,1.5) {\textbf{(a)}};
  \node[align=center,fill=none] at (1.5,1.5) {\textbf{(b)}};
  \end{tikzpicture}
  \caption{Reconstructed profiles of the flux pumping phase at 4.8s of the AUG discharge \#36663: (a) experimental (blue, with IMSE data) and modelled (red, without IMSE data) current densities; (b) corresponding effective electric field deficit. Reprinted from \cite{Burckhart2023NF}. © 2023 The Author(s). \href{https://creativecommons.org/licenses/by/4.0/}{CC BY 4.0}.}
\label{figAUGresults}
\end{figure}

Reproducible phases of flux pumping and sawteeth were obtained in AUG discharge \#36663 by adjusting the NBI (Neutral Beam Injection) heating power and the co-ECCD (Electron Cyclotron Current Drive) intensity \cite{Burckhart2023NF}. The sawtooth-free flux pumping state was specifically obtained at high normalized plasma beta ($\beta_N\sim 3$) and a moderate ECCD ($\sim 0.10\text{MA}$), in contrast to the two sawtoothing phases obtained at either lower $\beta_N$ ($\sim 2$) or higher ECCD ($\sim 0.15\text{MA}$), see Fig. 3 of \cite{Burckhart2023NF}. The deficits in the central current density ($\delta J_\varphi\simeq -2 \text{MA/m}^2$) and toroidal electric field ($\delta E_\varphi\simeq -8\text{mV/m}$) were identified by comparing the IDE (integrated data analysis equilibrium) solutions with and without considering the flux pumping mechanism [by including or excluding the Imaging Motional Stark Effect (IMSE) diagnostic data in IDE reconstructions], as shown by Fig. \ref{figAUGresults} (reprinted from Fig. 4 of \cite{Burckhart2023NF}). The experimental analyses demonstrate an anomalous radial redistribution of current density in the presence of m/n = 1/1 mode during the flux pumping phase.

\begin{figure}[htbp]
\centering
  \includegraphics[width=0.45\textwidth]{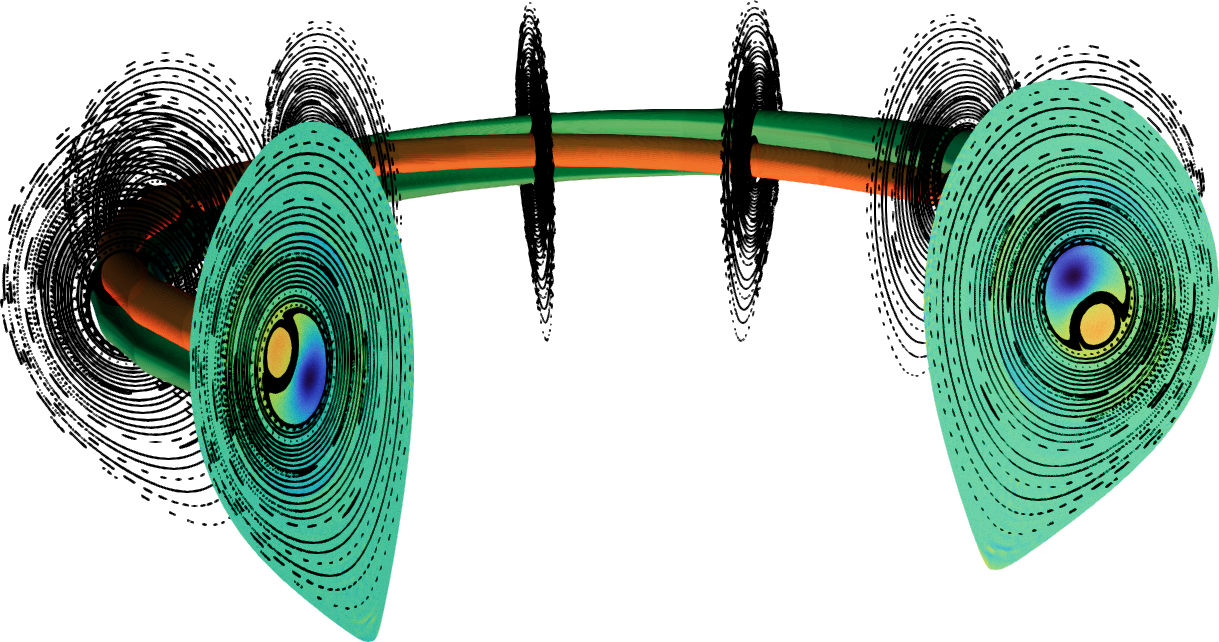}
  \caption{The quasi-stationary magnetic flux tubes in the 3D simulation of the AUG flux pumping discharge. The colored slices indicate the mode structure of toroidal magnetic field.}
\label{fig3Dstructure}
\end{figure}

To understand the self-regulating plasma dynamics during flux pumping, 3D quantitative nonlinear MHD simulations have been carried out at realistic parameters of the flux pumping phase of AUG discharge \#36663 ($q_0\simeq 1.0$, $\beta_N\simeq 3.0$, plasma current $I_p\simeq 0.8\text{MA}$, $\eta = 2.4\times10^{-9}\Omega\cdot\text{m}$, kinematic viscosity $\nu = 2.7\text{m}^2/\text{s}$) \cite{Burckhart2023NF, Zhang2025NFfluxpumping}. \revisioncolor{The simulations are carried out based on the two-temperature, single-fluid full MHD model of JOREK. The MHD equations and the experimental profiles employed in this study are presented in detail in \cite{Zhang2025NFfluxpumping}.} Compared with the published results of \cite{Zhang2025NFfluxpumping}, \revisioncolor{two} improvements have been \revisioncolor{introduced} to the simulation below by (i) introducing the equipartition terms in the temperature equations of ions and electrons for a more precise depiction of temperature and resistivity; (ii) applying isotropic resistivity in the Ohm's law with appropriate poloidal current sources instead of only in the toroidal direction. The 3D simulation presents a quasi-stationary helicity with two twisted flux tubes, as shown by Fig. \ref{fig3Dstructure}. The m/n = 1/1 mode dominates the 3D simulation across a resistive timescale of seconds, which is consistent with the experiments \cite{Burckhart2023NF}.

\begin{figure}[htbp]
\centering
  \begin{tikzpicture}
  \node[inner sep=0pt] (figs) at (0,0)
  {\includegraphics[width=0.33\textwidth]{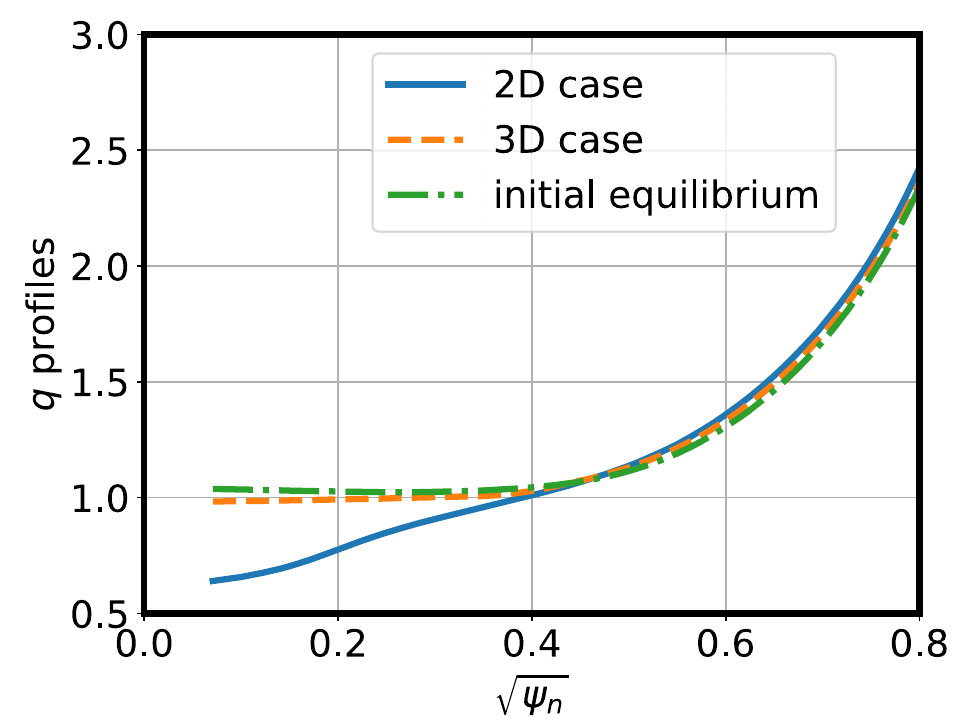}
  \includegraphics[width=0.33\textwidth]{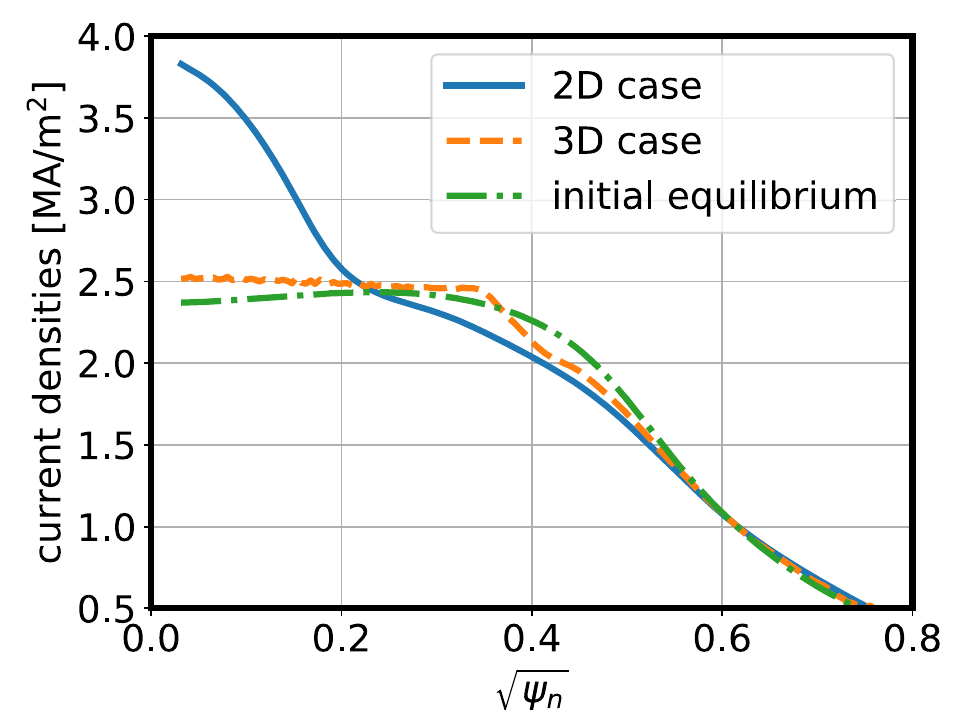}
  \includegraphics[width=0.32\textwidth]{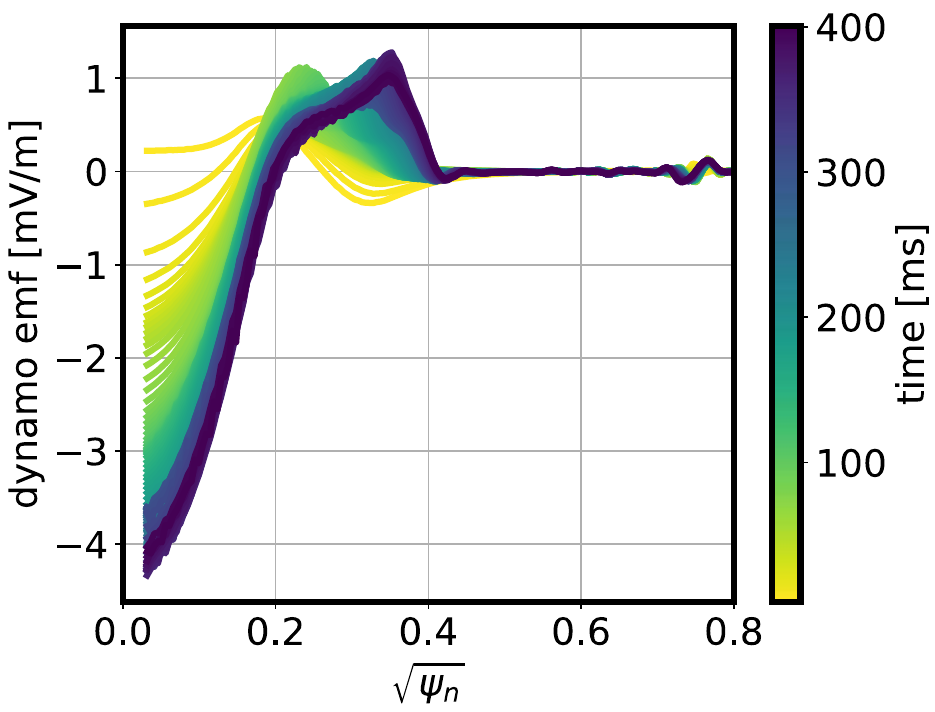}};
  \node[align=center,fill=none] at (-6.5,1.3) {\textbf{(a)}};
  \node[align=center,fill=none] at (-0.8,1.3) {\textbf{(b)}};
  \node[align=center,fill=none] at (4.0,1.3) {\textbf{(c)}};
  \end{tikzpicture}
  \caption{ (a) $q$ and (b) current density proﬁles at the saturated stages of the 2D (solid line) and 3D (dashed line) simulations, as well as of the initial equilibrium (dash-dotted line). (c) The parallel dynamo electric field vs. time from the 3D simulation.}
\label{fig2Dvs3D}
\end{figure}

A comparative analysis shown in Fig. \ref{fig2Dvs3D} was conducted between the 3D flux pumping modelling and the 2D current-diffusion modelling. The 2D case excludes non-axisymmetric MHD instabilities and focuses on the effect of the non-inductive current source on the 2D equilibrium evolution, with the pressure profile kept almost constant by balancing the heating source and heat conductivity induced diffusion. The non-inductive current source is the sum of NBI, ECCD, and bootstrap current, as shown in Fig. 19 (a) of \cite{Zhang2025NFfluxpumping}. The nonlinear 3D case, however, allows the self-consistent growth of n $\ge$ 1 MHD instabilities and the subsequent dynamo-induced redistribution of plasma profiles, such as current density and pressure. The saturated $q$ profile over the radial coordinate $\rho_p$ ($=\sqrt{\psi_n}$) from the 2D simulation is plotted by the solid line in Fig. \ref{fig2Dvs3D} (a), where $q_0$ decreases to 0.6 due to significant peak formation in the core current density. A value of $q_0\ll1$ is typically indicative of internal kink instability and sawtooth onset, which contradicts the experimental observation (sawtooth-free). In contrast, the 3D simulation with a dominant m/n = 1/1 mode shows that the saturated $q_0$ remains close to unity in the core, see the dashed line in Fig. \ref{fig2Dvs3D} (a). The comparison of toroidal current density in Fig. \ref{fig2Dvs3D} (b) confirms the discrepancy in $q$ profiles. In the 2D simulation, the central current density undergoes an increase from 2.4 MA/m$^2$ to 4 MA/m$^2$. However, in the 3D simulation, it is clamped around 2.5 MA/m$^2$, suggesting an effective redistribution of current density by the m/n = 1/1 mode induced flux pumping.

The current redistribution mainly results from the MHD dynamo generated by the m/n = 1/1 MHD instability \cite{Jardin2015PRL}, which exhibits characteristics of both quasi-interchange and tearing mode in the AUG modelling \cite{Zhang2025NFfluxpumping}. The dynamo electric field along the axisymmetric magnetic field ($\mathbf{b}_{0}$) is mainly generated by n = 1 non-axisymmetric components of plasma velocity $\Tilde{\mathbf{v}}$ and magnetic field $\Tilde{\mathbf{B}}$, i.e., $\varepsilon_\parallel=\langle(\Tilde{\mathbf{v}}\times\Tilde{\mathbf{B}})\cdot\mathbf{b}_{0}\rangle$, where $\langle\cdots\rangle$ denotes flux surface average. \revisioncolor{As shown in Fig. \ref{fig2Dvs3D} (c),} the dynamo generates a negative toroidal loop voltage of the order of mV/m in the core ($\rho_p < 0.2$), which equivalently increases the current diffusion. In the region beyond the plasma core ($0.2 < \rho_p < 0.4$), the dynamo loop voltage is positive, thereby increasing the current density. The reverse distribution of dynamo loop voltage continuously redistributes the current density and magnetic flux outward from the core region. As a result, the current density maintains a flat profile across a substantial radius ($\rho_p < 0.4$), and $q_0$ remains close to unity. The saturated $q$ profile, current density, and the dynamo electric field from the 3D simulation agree well with the experimental profiles reconstructed by including IMSE data during the flux pumping phase, as shown by Fig. \ref{figAUGresults} (reprinted from \cite{Burckhart2023NF}). 

\revisioncolor{The quantitative MHD modelling case of flux pumping as shown above was carried out with fully experimental parameters. Therefore, it will be treated as the base case in the study below, in which a broader parameter space will be scanned to extrapolate the parameter regimes of flux pumping.}
Compared with previous MHD modelling with lower Hartmann numbers ($10^3\lsim H\lsim 10^6$) \cite{Krebs2017PoPFluxPumping, Shen2018NF, Zhang2020NFfluxpumping, Cappello2000PRLRFP, Cappello2006PoPRFP, Bonfiglio2013PRL, Veranda2019NFRFX, Piovesan2017NF}, the above \revisioncolor{base case} demonstrates that the sawtooth-free quasi-stationary solution is also accessible at extremely low system dissipations \revisioncolor{(the Hartmann number $H_\text{base} \simeq 7.9\times10^7$, the magnetic Prandtl number $P_\text{base}\simeq 1.4\times10^3$, the Lundquist number $S_\text{base}\simeq 3\times10^{9}$, and the viscous Lundquist number $S_{\nu, base}\simeq 2\times10^{6}$)}. On this basis, the subsequent section systematically investigates the parameter regimes for bifurcations towards different plasma states, typically represented by flux pumping and sawteeth.

\section{Flux pumping and bifurcations of plasma core at different parameters}\label{Secparameterscans}
The full MHD equations used in this study are available in \cite{Zhang2025NFfluxpumping}. To facilitate the discussion, three characteristic times are defined: the resistive diffusion time $\tau_R=\mu_0L^2/\eta$, the viscous time $\tau_\nu=L^2/\nu$, and the Alfvén time $\tau_A=L/v_A$, where $\mu_0$ the vacuum magnetic permeability, $L$ the characteristic spatial scale, $\eta$ the resistivity, $\nu$ the kinematic viscosity, $v_A=B_0/\sqrt{\mu_0\rho_0}$ the Alfvén velocity, $B_0$ and $\rho_0$ the on-axis magnetic field strength and mass density. Following similar normalization procedure as \cite{Cappello2006PoPRFP}: $t^\prime=t\sqrt{\tau_\nu/\tau_R}/\tau_A$, $\mathbf{v}^\prime=\mathbf{v}\sqrt{\tau_R/\tau_\nu}/v_A$, $\rho^\prime=\rho/\rho_0$, $\nabla^\prime=L\nabla$, $\mathbf{B}^\prime=\mathbf{B}/B_0$, $p^\prime=\mu_0p/B_0^2$, $\mathbf{J}^\prime=\mu_0\mathbf{J}L/B_0$, $\mathbf{S}_j^\prime=\mu_0\mathbf{S}_jL/B_0$, the induction equation and momentum equation can be rewritten as (ignore $^\prime$ for all normalized quantities)
\begin{equation}\label{eq1induction}
  \begin{split}
    \dfrac{\partial\mathbf{B}}{\partial t}=\nabla\times\left(\mathbf{v}\times\mathbf{B}\right)-\nabla\times\left[H^{-1}\left(\nabla\times\mathbf{B}-\mathbf{S}_j\right)\right],
  \end{split}
\end{equation}
\begin{equation}\label{eq2momentum}
  \begin{split}
    P^{-1}\left(\dfrac{\partial\mathbf{v}}{\partial t}+\mathbf{v}\cdot\nabla\mathbf{v}\right)=\rho^{-1}\left[\left(\nabla\times\mathbf{B}\right)\times\mathbf{B}-\nabla p\right]+\nabla\cdot\left(H^{-1}\nabla\mathbf{v}\right).
  \end{split}
\end{equation}
The adopted full MHD model in JOREK exhibits slight discrepancies from the prior dual-field MHD model (only solving for $\mathbf{v}$ and $\mathbf{B}$) in RFP modelling \cite{Cappello2000PRLRFP, Cappello2006PoPRFP}: (i) the pressure gradient enters the system and introduces an additional degree of freedom in the driving intensity for the mode; (ii) the density is not fixed but it is expected to be stable in nonlinear 3D simulations. It is found that beyond the continuity and energy equations, the nonlinear MHD system is mainly governed by two dimensionless parameters $H$ and $P$ \cite{Cappello2000PRLRFP, Cappello2006PoPRFP}, where the Hartmann number $H\equiv \sqrt{\tau_R\tau_\nu}/\tau_A=\sqrt{\mu_0/(\eta\nu)}Lv_A$, and the magnetic Prandtl number $P\equiv\tau_R/\tau_\nu=\mu_0\nu/\eta$. For specific cases with negligible inertia terms on the left-hand side, $H$ is the unique control parameter for the plasma state, which is inversely proportional to the system dissipation. From Eq. \ref{eq1induction}, we can also infer that a smaller Hartmann number (larger system dissipation) can generate stronger MHD instabilities to modulate the magnetic field and the relevant $q$ profile through the nonlinear dynamo term, i.e., $\Tilde{\mathbf{v}}\times\Tilde{\mathbf{B}}\sim H^{-1}(\mathbf{J}_0-\mathbf{S}_j)$. Therefore, to facilitate the comparison of dynamos at different Hartmann numbers, the normalized dynamo term extracted from Eq. \ref{eq1induction} will be further normalized by $(H/H_\text{base}\cdot\sqrt{P_\text{base}}/v_A)^{-1}$ below, unless specified otherwise. Note that the constant $(\sqrt{P_\text{base}}/v_A)^{-1}$ originates from the initial normalization of $\mathbf{v}$ and is introduced to restore the amplitude of the normalized dynamo of the base case to its SI unit value, i.e., of the order of mV/m. \revisioncolor{The simulations in this work were carried out with (100, 100) grids in poloidal and radial directions, incorporating toroidal harmonics ranging from 0 to 4.}

\subsection{Bifurcated plasma states: flux pumping, sawtooth, and others.}

\begin{figure}[htbp]
\centering
  \includegraphics[width=0.6\textwidth]{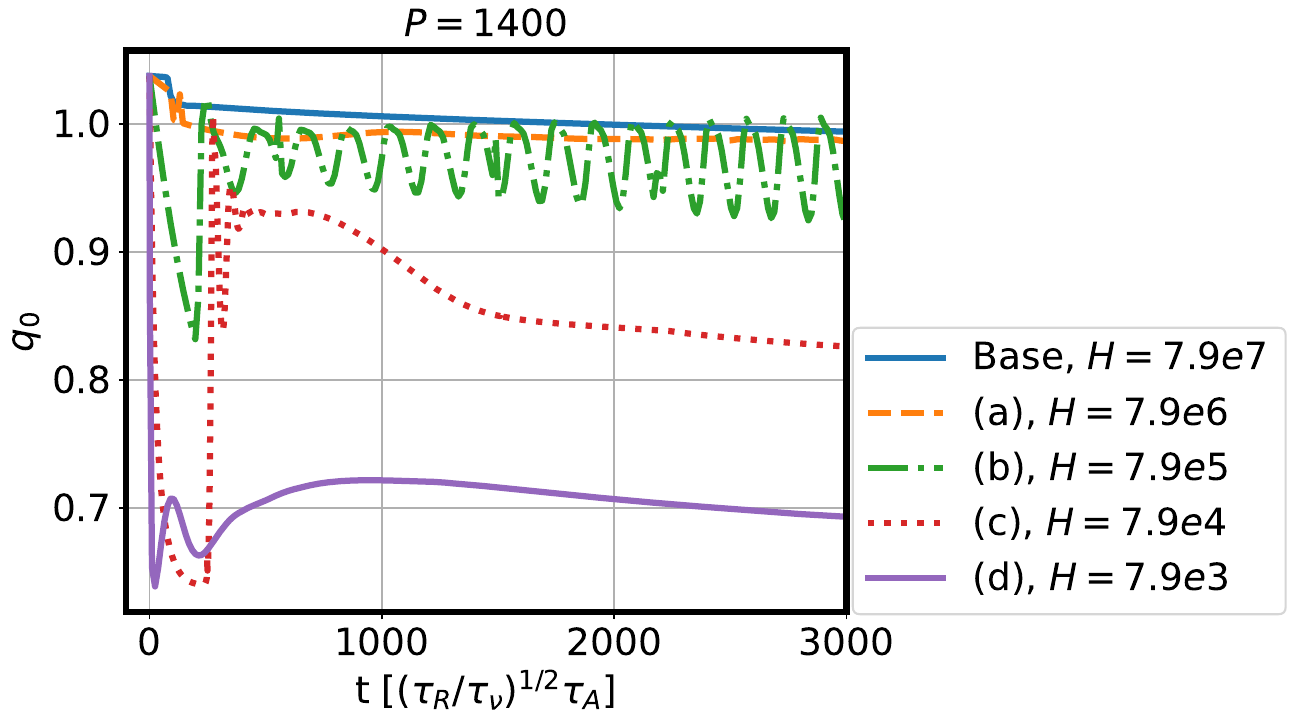}
  \caption{Different plasma states obtained at different system dissipations: Base case with $H=7.9\times10^7$ (blue solid line); case (a) with $H=7.9\times10^6$ (orange dashed line); case (b) with $H=7.9\times10^5$ (green dash-dotted line); case (c) with $H=7.9\times10^4$ (red dotted line); and case (d) with $H=7.9\times10^3$ (purple solid line). $P$ is fixed at 1400 for all cases.}
\label{figfourstates}
% [00_etax10_viscox10_Hx0.1_Px1]# plot_multiple_files.py ../Base_case/postproc/q_vs_time_at_rhop_eq_0.07.dat postproc/q_vs_time_at_rhop_eq_0.07.dat ../01_etax100_viscox100_Hx0.01_Px1/postproc/q_vs_time_at_rhop_eq_0.07.dat ../02_etax1000_viscox1000_Hx0.001_Px1/postproc/q_vs_time_at_rhop_eq_0.07.dat ../03_etax10000_viscox10000_Hx0.0001_Px1/postproc/q_vs_time_at_rhop_eq_0.07.dat -xm 57.8368999421631 -xl 't [$(\tau_R/\tau_\nu)^{1/2}\tau_A$]' -yl '$q_0$' -legends 'Base, $H = 7.9e7$' '(a), $H = 7.9e6$' '(b), $H = 7.9e5$' '(c), $H = 7.9e4$' '(d), $H = 7.9e3$' -ti '$P=1400$' -xlim -100 2000
% plot_multiple_files.py ../Base_case/postproc/q_vs_time_at_rhop_eq_0.07.dat postproc/q_vs_time_at_rhop_eq_0.07.dat ../01_etax100_viscox100_Hx0.01_Px1/postproc/q_vs_time_at_rhop_eq_0.07.dat ../02_etax1000_viscox1000_Hx0.001_Px1_with_densitysource/postproc/q_vs_time_at_rhop_eq_0.07.dat ../03_etax10000_viscox10000_Hx0.0001_Px1/postproc/q_vs_time_at_rhop_eq_0.07.dat -xm 57.8368999421631 -xl 't [$(\tau_R/\tau_\nu)^{1/2}\tau_A$]' -yl '$q_0$' -legends 'Base, $H = 7.9e7$' '(a), $H = 7.9e6$' '(b), $H = 7.9e5$' '(c), $H = 7.9e4$' '(d), $H = 7.9e3$' -ti '$P=1400$' -xlim -100 3000
\end{figure}

First, to evaluate the influence of the overall system dissipation (represented by the reciprocal of Hartmann number) on the core plasma state, the ratio of $\nu/\eta$ and all equilibrium parameters are fixed same as the base case ($\eta_\text{base}=2.4\times10^{-9}\Omega\cdot\text{m}$, $\nu_\text{base}=2.7\text{m}^2/\text{s}$, $P_\text{base}=1400$, $\beta_N\simeq3$) in Sec. \ref{Sec3DBaseCase}. Then, $\nu$ and $\eta$ are increased simultaneously to decrease $H$. Four representative cases are simulated respectively at the increased $\eta$ and $\nu$ by (a) $10\times$, (b) $10^2\times$, (c) $10^3\times$, and (d) $10^4\times$, corresponding to decreasing $H$ by (a) $10^{-1}\times$, (b) $10^{-2}\times$, (c) $10^{-3}\times$, and (d) $10^{-4}\times$. Four different plasma states are obtained at varying dissipation levels, as shown by the temporal evolutions of $q_0$ in Fig. \ref{figfourstates}. For the two low dissipation cases [the base case and case (a), $H=7.9\times10^7$ and $\times10^6$], the quasi-stationary flux pumping solution is obtained, and $q_0$ remains close to unity throughout. However, as the dissipation is further increased by one order of magnitude [case (b), $H=7.9\times10^5$], the plasma core exhibits the sawtooth-like kink cycling state with $q_0$ oscillating between 0.93 and 1.0. The oscillations of pressure and current density on the axis are also observed (not shown). Furthermore, when the dissipation is increased to reach $H=7.9\times10^4$ in case (c), $q_0$ first decreases below 0.7 and then triggers a giant sawtooth crash ($q_0$ recovers to unity quickly). Afterwards, $q_0$ is sustained below unity for thousands of normalized time units until the simulation terminates. The late stage is analogous to the saturated m/n = 1/1 resistive internal kink mode. For the last case (d) with the highest dissipation and $H=7.9\times10^3$, $q_0$ decreases dramatically in a short timescale \revisioncolor{below 100 normalized time units} and then saturates around 0.7 for thousands of normalized time units. The $q_0$ evolution of case (d) is similar to the 2D case, though a m/n = 1/1 magnetic island is observed in the plasma core, suggesting that the m/n = 1/1 mode may not affect the $q$ profile here. The aforementioned cases are categorized into four types based on the distinct characteristics of $q_0$ evolution: (i) the flux pumping state (FP), (ii) the sawtooth state (ST), (iii) the single-crash state (SC) followed by a quasi-stationary m/n = 1/1 magnetic island, and (iv) the quasi-stationary m/n = 1/1 magnetic island state (QS), respectively.

\begin{figure}[htbp]
\centering
  \begin{tikzpicture}
  \node[inner sep=0pt] (figs) at (0,0)
  {\includegraphics[width=0.5\textwidth]{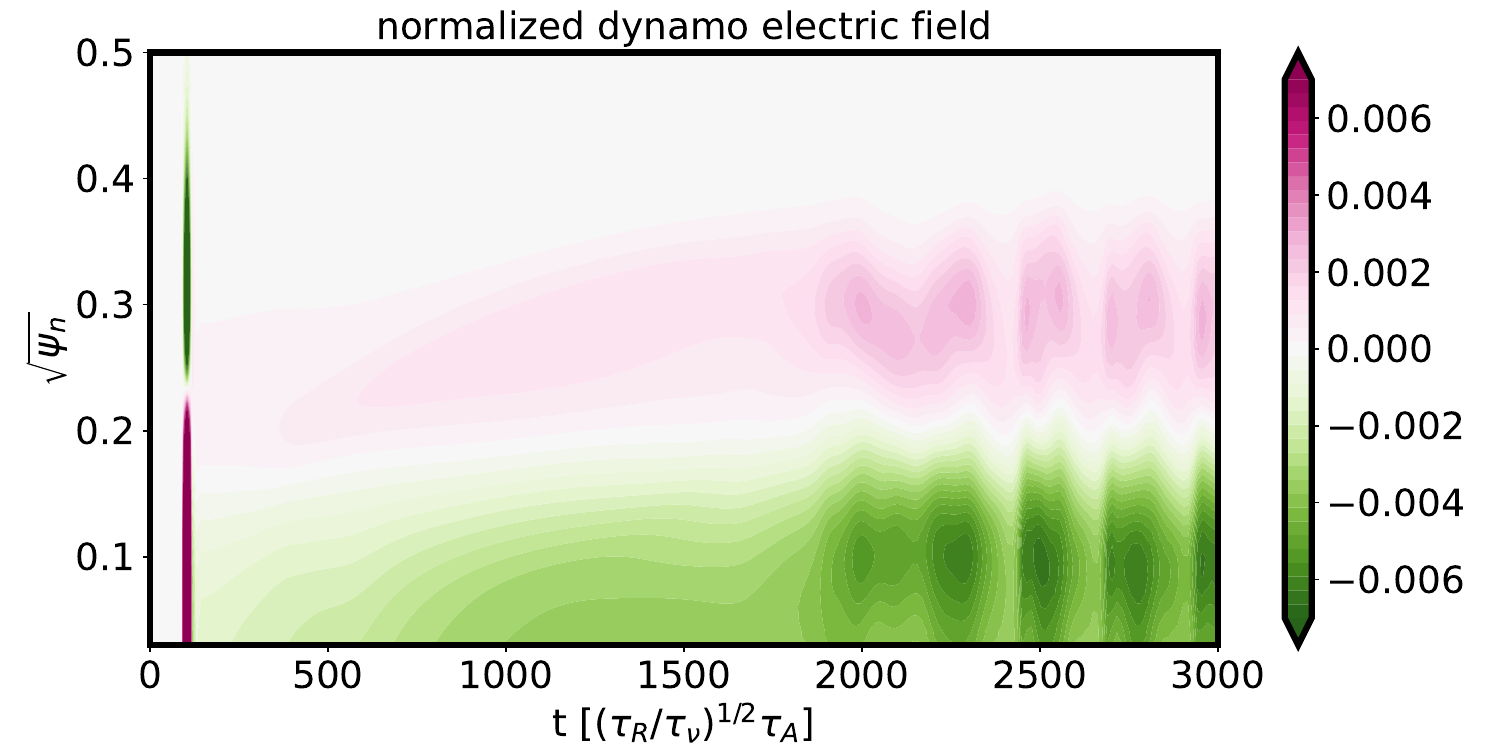}
  \includegraphics[width=0.5\textwidth]{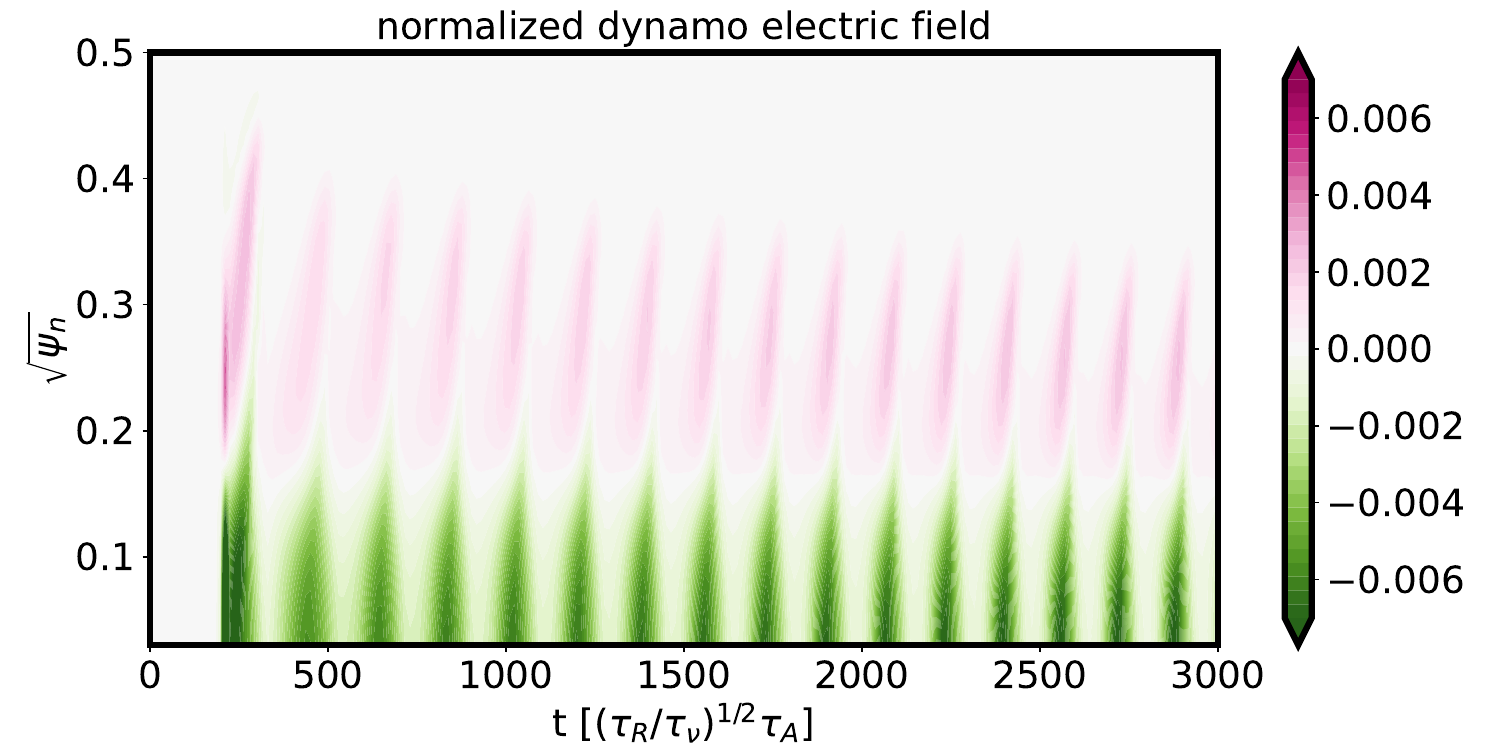}};
  \node[inner sep=0pt] (figs) at (0,-4.2)
  {\includegraphics[width=0.5\textwidth]{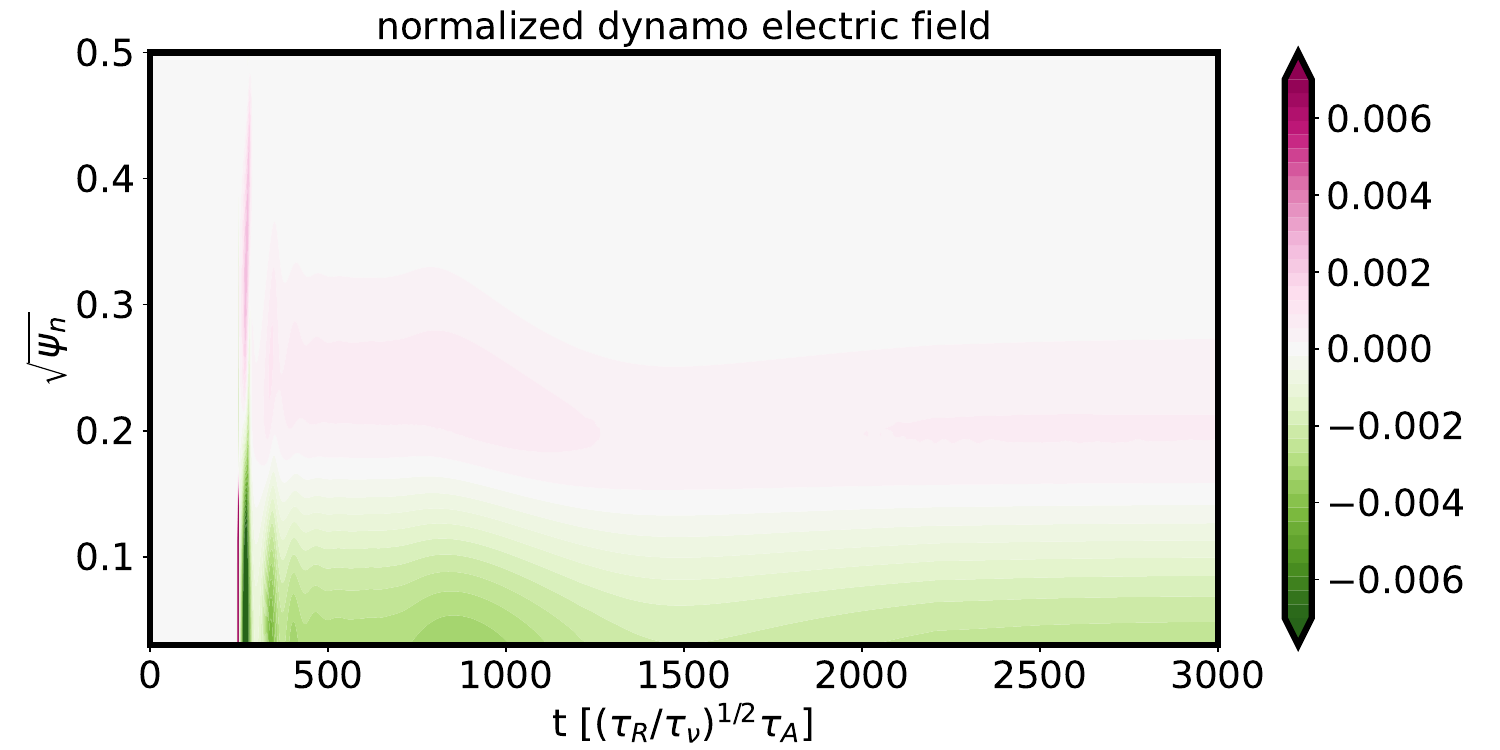}
  \includegraphics[width=0.5\textwidth]{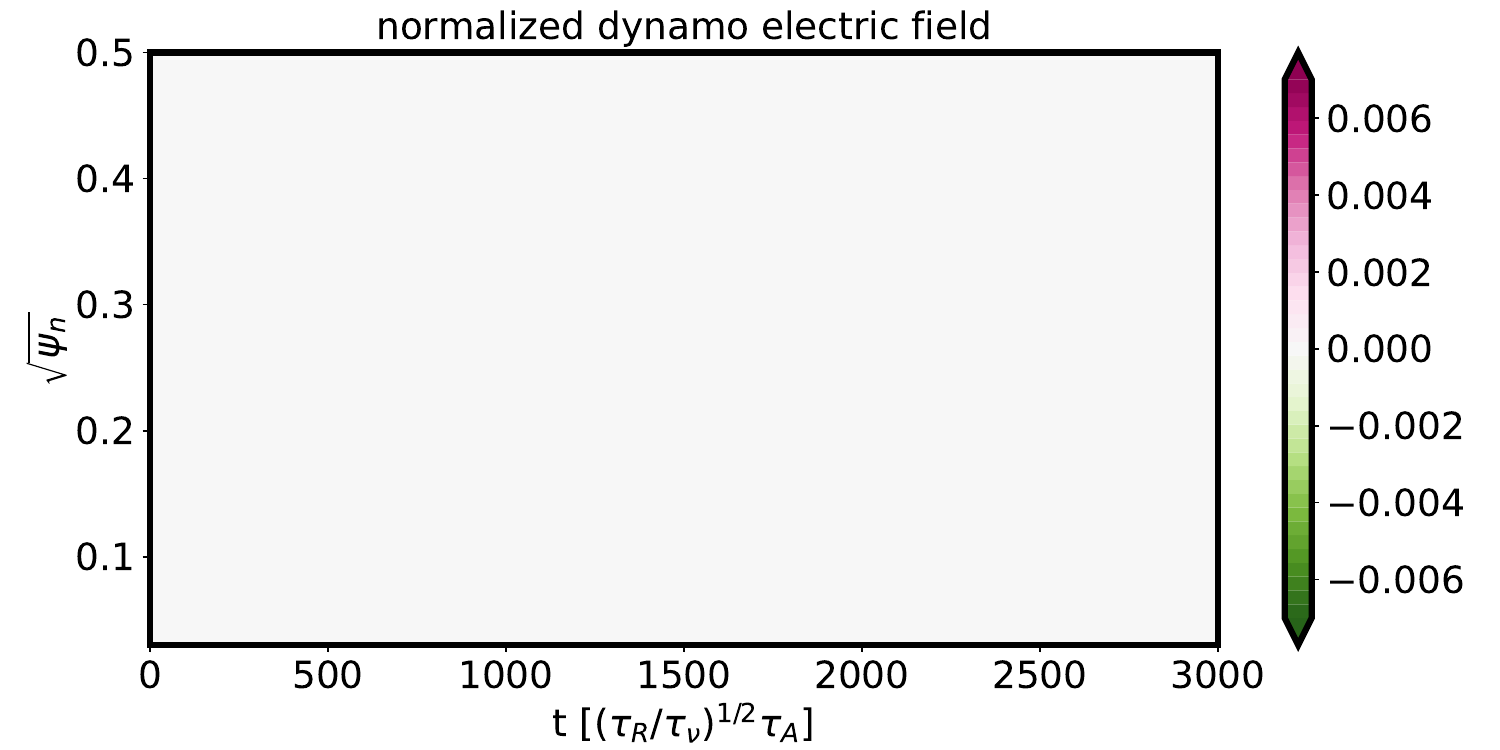}};
  \node[align=center,fill=none] at (-6.5, 1) {\textbf{(a)}};
  \node[align=center,fill=none] at ( 1.5, 1) {\textbf{(b)}};
  \node[align=center,fill=none] at (-6.5,-3.3) {\textbf{(c)}};
  \node[align=center,fill=none] at ( 1.5,-3.3) {\textbf{(d)}};
  \end{tikzpicture}
  \caption{Temporal evolutions of the radial profile of dynamo electric field normalized by $(H/H_\text{base}\cdot\sqrt{P_\text{base}}/v_A)^{-1}$, respectively for cases (a) $H=7.9\times10^6$, (b) $H=7.9\times10^5$, (c) $H=7.9\times10^4$, and (d) $H=7.9\times10^3$. $P$ is fixed at 1400 for all cases.}
\label{figdynamofourcases}

\end{figure}

\begin{figure}[htbp]
\centering
  \begin{tikzpicture}
  \node[inner sep=0pt] (figs) at (0,0)
  {\includegraphics[width=0.4\textwidth]{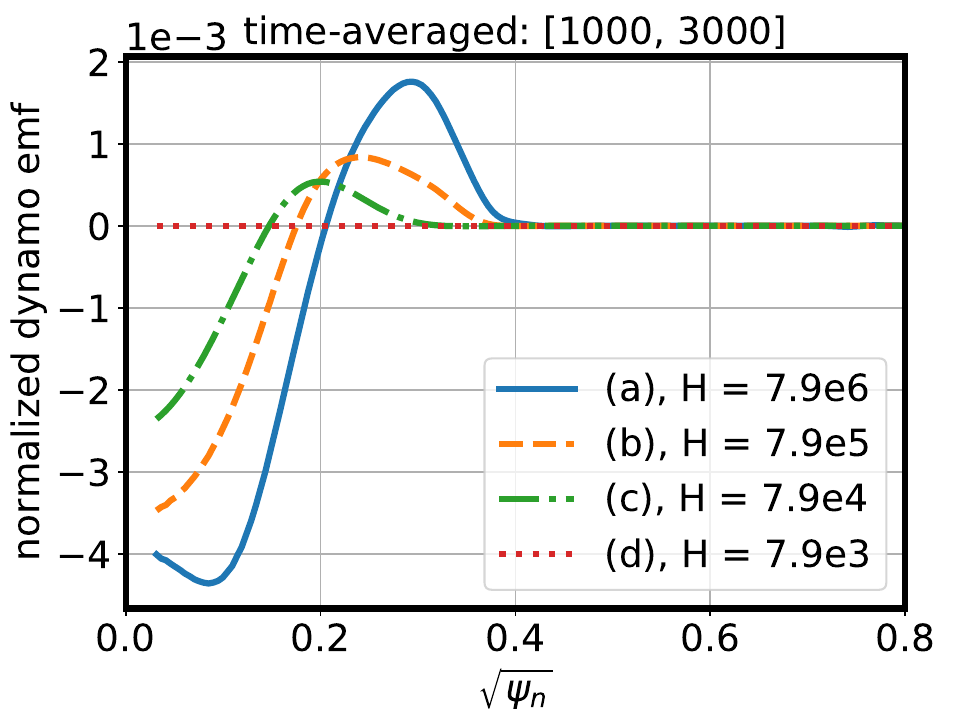}
  \includegraphics[width=0.4\textwidth]{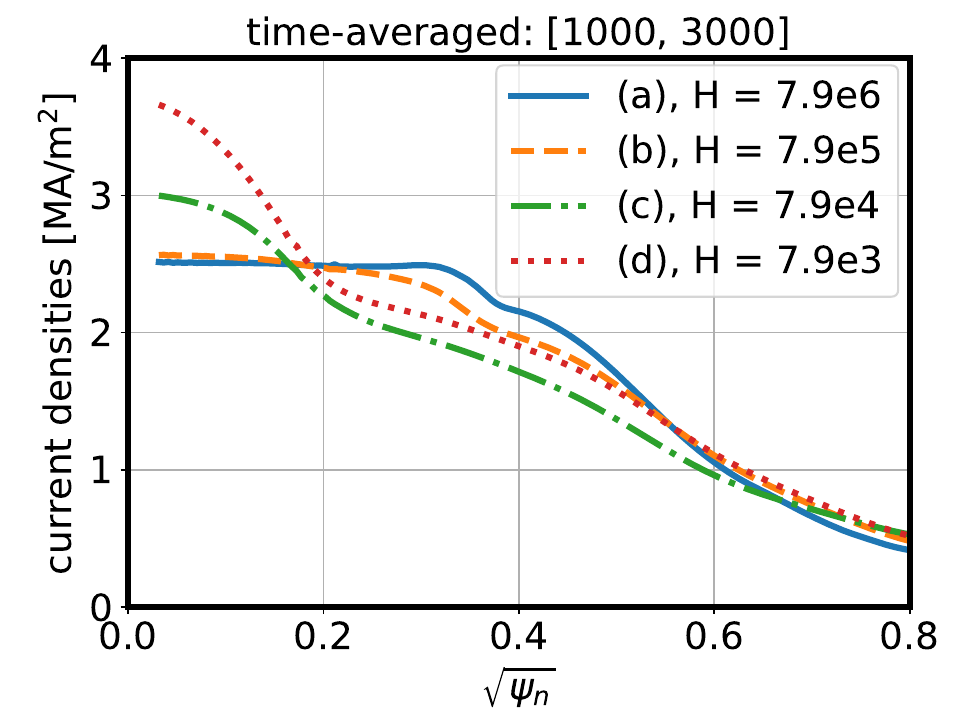}};
  \node[align=center,fill=none] at (-5, 1.7) {\textbf{(a)}};
  \node[align=center,fill=none] at ( 1.7, 1.7) {\textbf{(b)}};
  \end{tikzpicture}
  \caption{\revisioncolor{The time-averaged radial profiles of the dynamo electric field [normalized by $(H/H_\text{base}\cdot\sqrt{P_\text{base}}/v_A)^{-1}$] and current density, respectively, for four cases: solid line, case (a) with $H=7.9\times10^6$; dashed line, case (b) with $H=7.9\times10^5$; dash-dotted line, case (c) with $H=7.9\times10^4$; and dotted line, case (d) with $H=7.9\times10^3$. $P$ is fixed at 1400 for all cases. The time interval used to calculate the average is [1000, 3000].}}
\label{figdynamo_Jtor_avg}

% plot_multiple_files 00_etax10_viscox10_Hx0.1_Px1/postproc/avg_VXB_para_profile_during_t_eq_1000.00_3000.00.dat 01_etax100_viscox100_Hx0.01_Px1/postproc/avg_VXB_para_profile_during_t_eq_1000.00_3000.00.dat 02_etax1000_viscox1000_Hx0.001_Px1_with_densitysource/postproc/avg_VXB_para_profile_during_t_eq_1000.00_3000.00.dat 03_etax10000_viscox10000_Hx0.0001_Px1/postproc/avg_VXB_para_profile_during_t_eq_1000.00_3000.00.dat -xl '$\sqrt{\psi_n}$' -yl 'normalized dynamo emf' -legends '(a), H = 7.9e6' '(b), H = 7.9e5' '(c), H = 7.9e4' '(d), H = 7.9e3' -xlim 0 0.8 -sk 0 -ti 'time-averaged: [1000, 3000]'

% plot_multiple_files 00_etax10_viscox10_Hx0.1_Px1/postproc/avg_Jtor_profile_during_t_eq_1000.00_3000.00.dat 01_etax100_viscox100_Hx0.01_Px1/postproc/avg_Jtor_profile_during_t_eq_1000.00_3000.00.dat 02_etax1000_viscox1000_Hx0.001_Px1_with_densitysource/postproc/avg_Jtor_profile_during_t_eq_1000.00_3000.00.dat 03_etax10000_viscox10000_Hx0.0001_Px1/postproc/avg_Jtor_profile_during_t_eq_1000.00_3000.00.dat -xl '$\sqrt{\psi_n}$' -yl 'current densities [MA/m$^2$]' -legends '(a), H = 7.9e6' '(b), H = 7.9e5' '(c), H = 7.9e4' '(d), H = 7.9e3' -ym '1e-6' -ylim 0 4 -xlim 0 0.8 -sk 0 -ti 'time-averaged: [1000, 3000]'

\end{figure}

The temporal evolutions of the normalized dynamo electric field [by $(H/H_\text{base}\cdot\sqrt{P_\text{base}}/v_A)^{-1}$, as described above] for cases (a-d) are plotted in Fig. \ref{figdynamofourcases}. \revisioncolor{Note that the dynamo evolution of the base case is similar to case (a) and is omitted here, which can be found in Fig. \ref{fig2Dvs3D} (c) and in Fig. 9 of \cite{Zhang2025NFfluxpumping}.} In all three cases (a-c) with prominent dynamo activities, the dynamo profile is similar to Fig. \ref{fig2Dvs3D} (c). In the flux pumping case shown by Fig. \ref{figdynamofourcases} (a), the dynamo remains stable around the normalized amplitude of $0.004\sim0.007$ after $t=1000$, although a limited fluctuation is observed after $t=2000$. The stable dynamo is responsible for maintaining the saturated $q_0\simeq1.0$ in the flux pumping case by redistributing the applied non-inductive current source. As will be discussed in Sec. \ref{subsecBreakDownofFP}, the fluctuation of the dynamo in Fig. \ref{figdynamofourcases} (a) is caused by the destabilization of n $\ge$ 2 harmonics, which reduce the robustness of flux pumping dynamo and eventually lead to the \revisioncolor{sawtooth onset} in long-term simulations. In contrast, the dynamo in the sawtoothing case (b) oscillates at the same frequency as $q_0$, as shown by Fig. \ref{figdynamofourcases} (b). The maximum normalized dynamo amplitude during the sustained sawtooth cycles is about 0.007, which flattens the current density on a timescale much shorter than the resistive diffusion, thereby appearing mainly at the rising phase of $q_0$. After the m/n = 1/1 magnetic island replaces the original magnetic axis and restores the axisymmetric flux surfaces, the dynamo amplitude decreases to a negligible magnitude due to the decay of the pumping m/n = 1/1 resistive internal kink mode. For case (c) with the single crash in Fig. \ref{figdynamofourcases} (c), a strong dynamo of the normalized magnitude of 0.012 is observed at $t=270$, lifting $q_0$ from 0.65 to unity on a short timescale of $\sim 20 (\tau_R/\tau_\nu)^{1/2}\tau_A$ ($\sim 0.34$ ms). Afterwards, the normalized amplitude of dynamo saturates around 0.003 and maintains $q_0$ around 0.9 without further crashes in the simulated timescale. For case (d) with a quasi-stationary m/n = 1/1 magnetic island as depicted in Fig. \ref{figdynamofourcases} (d), the normalized dynamo amplitude is at the noise level and not able to counteract the current source anymore. Therefore, the temporal evolution of \revisioncolor{the} $q$ profile resembles the 2D case with $q_0\ll 1.0$.

\revisioncolor{Fig. \ref{figdynamo_Jtor_avg} illustrates the time-averaged radial profiles of the normalized dynamo electric field and the current density for the above four cases. Consistent with Fig. \ref{figdynamofourcases}, in the inner region ($\sqrt{\psi_n} < 0.4$), higher Hartmann numbers drive a larger and broader dynamo electric field, whereas the lowest $H$ case (dotted line) shows near-total suppression of the dynamo electric field. This variation is mirrored in the current density profiles, where lower Hartmann numbers result in a more peaked distribution. However, as $H$ increases, the current profiles tend to converge to the flat profile, particularly in the core region. The time-averaged dynamo electric fields for cases (a)-(c) have comparable amplitudes, which indicates the effective role played by the dynamo in redistributing current density and magnetic flux, either through quasi-stationary flux pumping or periodic sawtooth relaxations.}

\subsection{From flux pumping to sawtooth at varying Hartmann numbers}

\begin{figure}[htbp]
  \centering
  \begin{tikzpicture}
    \node[inner sep=0pt] (figs) at (0,0)
    {\includegraphics[width=\textwidth]{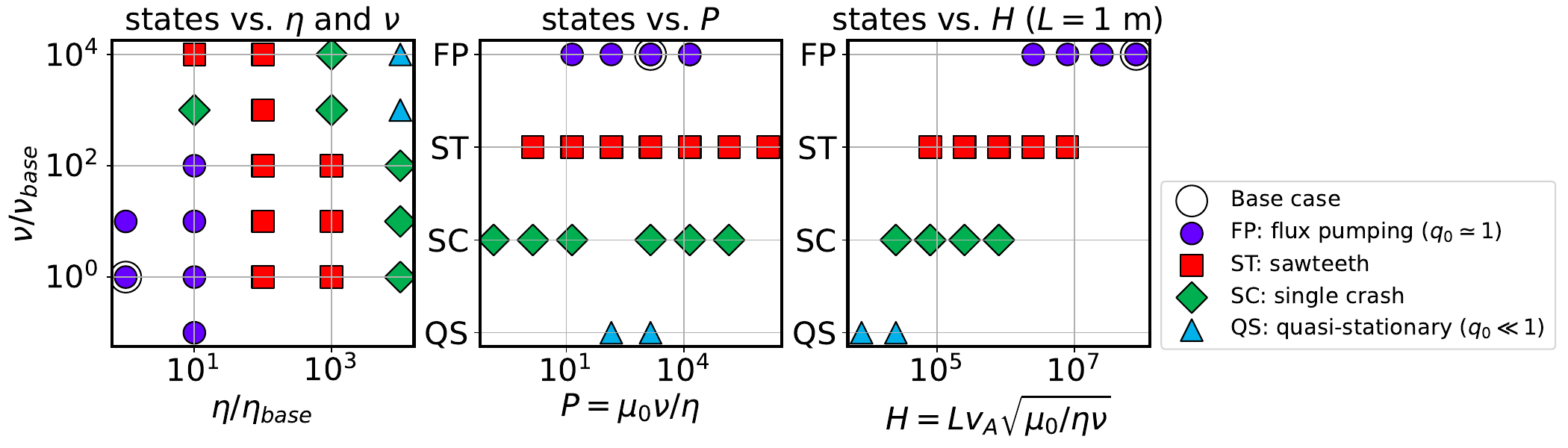}};
    \node[align=center,fill=none] at (-6.4,1.2) {\textbf{(a)}};
    \node[align=center,fill=none] at (-2.55,1.2) {\textbf{(b)}};
    \node[align=center,fill=none] at ( 1.1,1.2) {\textbf{(c)}};
  \end{tikzpicture}
    \caption{Different plasma states obtained by scanning viscosity and resistivity over a broad 2D parameter space from $1\times\eta_\text{base}$ to $10^4\times\eta_\text{base}$ and from $0.1\times\nu_\text{base}$ to $10^4\times\nu_\text{base}$ (a). The dependencies of plasma states are plotted against the magnetic Prandtl number (b) and the Hartmann number (c).}
  \label{figureStatesAllscans}
\end{figure}
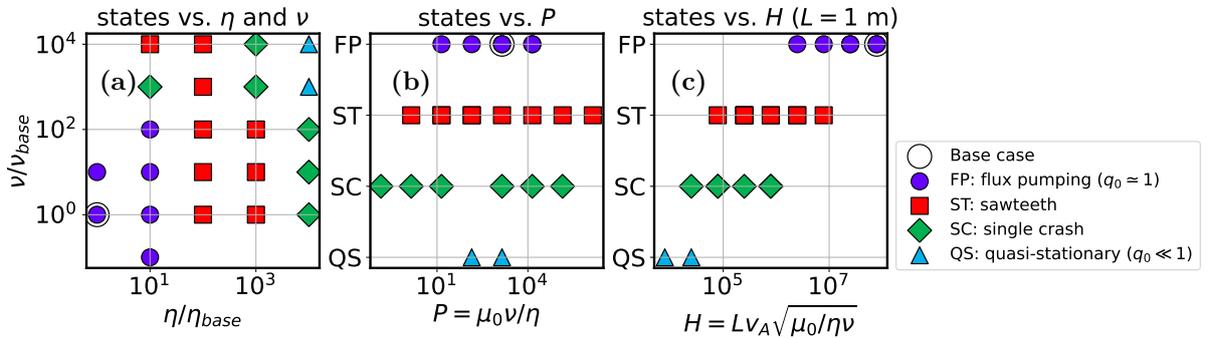

The viscosity and resistivity of interest are further extrapolated to a broad 2D map from $1\times\eta_\text{base}$ to $10^4\times\eta_\text{base}$ and from $0.1\times\nu_\text{base}$ to $10^4\times\nu_\text{base}$ at $\beta_N\simeq3$, while other parameters remain unchanged. The plasma state for each set of viscosity and resistivity is illustrated in Fig. \ref{figureStatesAllscans} (a). The diagram shows that flux pumping can be obtained at relatively low dissipation, as shown by the cycles in the lower left corner of Fig. \ref{figureStatesAllscans} (a). Conversely, the sawtooth and single-crash states with sawtooth relaxation are obtained at moderate dissipation (rectangles and diamonds). The quasi-stationary island state is only accessible at extremely high dissipation (up triangles). Meanwhile, there is no clear dependence of plasma states on the magnetic Prandtl number in this tokamak modelling, as evidenced by Fig. \ref{figureStatesAllscans} (b), which is qualitatively consistent with the RFP modelling \cite{Cappello2000PRLRFP, Cappello2006PoPRFP}. Instead, the transitions of plasma states can be more intuitively represented by the Hartmann number, which acts as a reverse proxy for system dissipation. As shown by Fig. \ref{figureStatesAllscans} (c), from low Hartmann number $H\sim10^3$ to high Hartmann number $H\sim10^8$, the plasma state transitions from quasi-stationary island, single-crash state, sawtooth, and flux pumping in sequence. The strong dependence of central plasma states on the Hartmann number resembles that in RFP modelling. \revisioncolor{Though the laminar QS/QSH state was obtained at low Hartmann number ($H\lsim 10^3$) in the early modelling of RFX-mod \cite{Cappello2000PRLRFP, Cappello2006PoPRFP}, recent modelling with helical boundary conditions has also reproduced the QS/QSH state at experimentally relevant regimes ($H\gsim10^5$) \cite{Bonfiglio2013PRL, Veranda2019NFRFX}.}

\revisioncolor{Similar to Fig. \ref{figdynamo_Jtor_avg}, for the most cases of Fig. \ref{figureStatesAllscans} with relatively low viscosities ($\nu/\nu_\text{base}\le10^2$), the magnitude of the time-averaged normalized dynamo remains roughly stable (of the order of $\sim10^{-3}$) as the Hartmann number changes (not shown here), thereby maintaining $q_0$ much larger than in the 2D case ($q_0\sim0.6$). Nevertheless, there are two distinct points represented by the blue triangles in Fig. \ref{figureStatesAllscans} at $H<10^{-3}H_\text{base}\simeq7.9\times10^4$, in which the normalized dynamo amplitude dramatically falls below the average ($\sim10^{-3}$), as indicated by Fig. \ref{figdynamo_Jtor_avg}.} These two cases correspond to the quasi-stationary island state in Figs. \ref{figfourstates}-\ref{figureStatesAllscans}. The absence of a dynamo electric field results in the peaking of central current density and the drop in $q_0$, which are more analogous to the 2D situation.

\begin{figure}[htbp]
  \centering
    \includegraphics[width=0.5\textwidth]{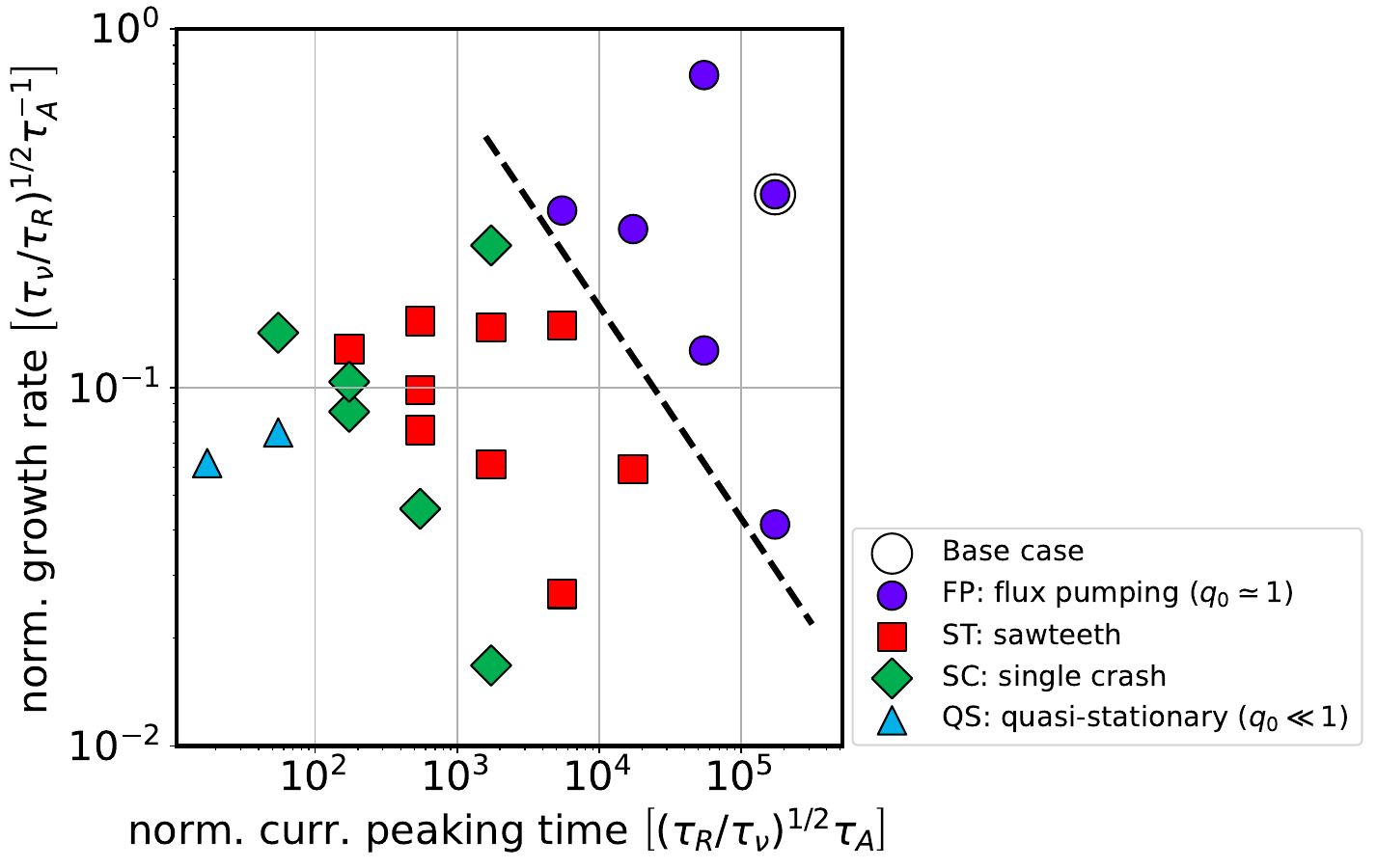}
    \caption{The plasma states classified by the linear growth rate of m/n = 1/1 (resistive) quasi-interchange mode in 3D simulations and the current peaking times in 2D simulations. The growth rate and current peaking time are normalized with the same method as Eqs. \ref{eq1induction} and \ref{eq2momentum}. The dashed line roughly marks the boundary for accessing flux pumping.}
  \label{figureLinearGRandCurrentPeakingTime}
\end{figure}

3D nonlinear MHD modelling of flux pumping remains computationally time-consuming. The surrogate model being developed aims to efficiently predict flux-pumping accessibility using a flight simulator \cite{Krebs2023EPSreduced}. Prior to this, the underlying relationship between the linear growth rate of the m/n = 1/1 (resistive) quasi-interchange mode in 3D simulations and the current peaking timescale in 2D simulations is investigated. The growth rate and current peaking timescales are normalized following the same procedure for Eqs. \ref{eq1induction} and \ref{eq2momentum}. As shown in Fig. \ref{figureLinearGRandCurrentPeakingTime}, flux pumping is mainly accessible when the normalized linear growth rate of the mode is sufficiently large ($\gsim 10^{-1}$) and the normalized characteristic current peaking timescale is long enough ($\gsim 10^4$). Otherwise, states with sawtooth relaxations are likely to be retained. The primary benefits of bridging plasma states with the linear growth rate and current peaking timescale are largely attributed to the reduced requirement of computational resources, as only the 2D modelling and linear phase of 3D cases are needed. The larger \revisioncolor{normalized} growth rate potentially indicates a stronger nonlinear dynamo term generated by the instability, and the longer current peaking timescale corresponds to a lower current driving efficiency, thereby favoring flux pumping. Nevertheless, more in-depth analyses are necessary and are being carried out to build the knowledge to accurately assess the feasibility of flux pumping without full-cycle 3D nonlinear MHD modelling.

\subsection{Breakdown of flux pumping and sawtooth onset at marginal beta and dissipation} \label{subsecBreakDownofFP}

\begin{figure}[h]
  \centering
    \includegraphics[width=0.7\textwidth]{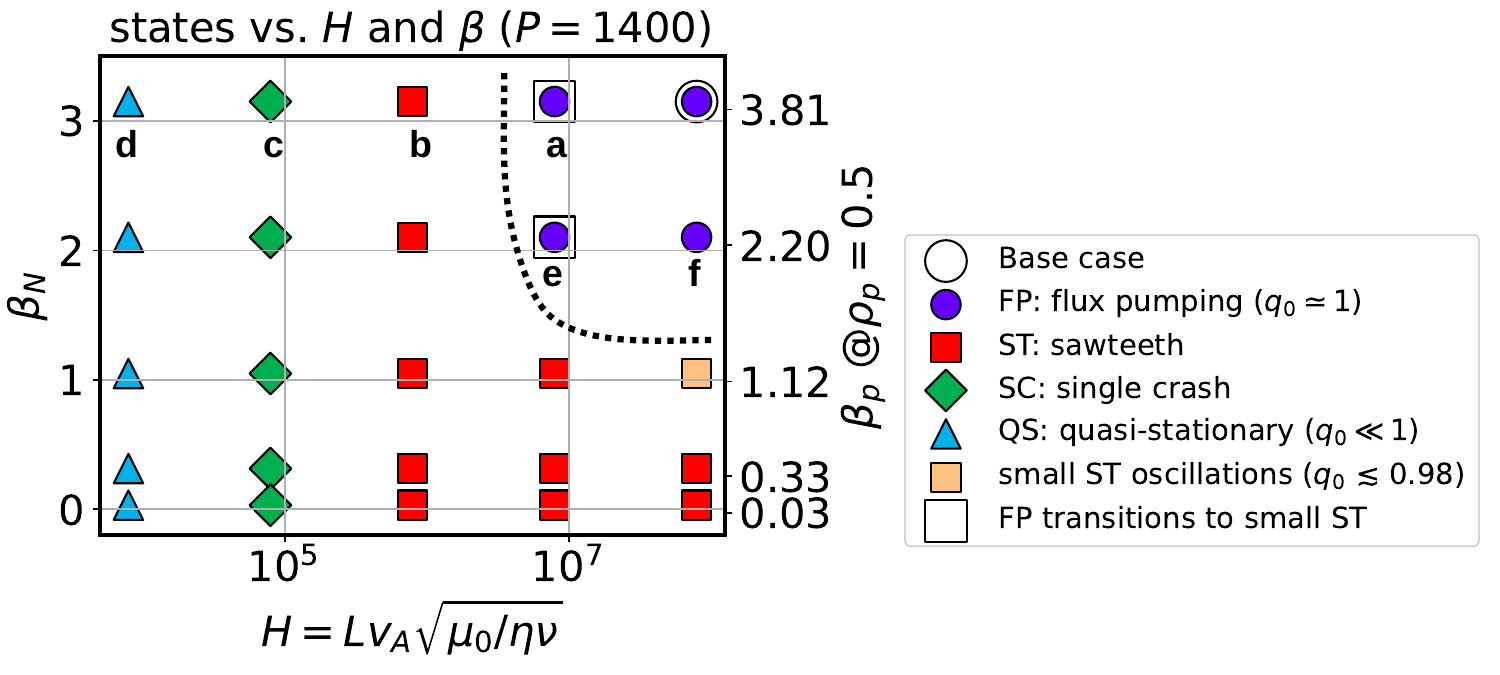}
    \caption{Different plasma states at different Hartmann numbers and plasma betas. $\beta_p$ is evaluated at half the minor radius. The dashed line roughly marks the boundary for accessing flux pumping. $P$ is fixed at 1400 for all cases. The four states in the Fig. \ref{figfourstates} correspond to cases (a-d) with $\beta_N\simeq3$. Transitions from flux pumping to small sawtooth are observed in cases (a) and (e) in longer timescale simulations ($\sim1.5\times10^4$ $(\tau_R/\tau_\nu)^{1/2}\tau_A\sim 260$ ms). Cases (e) and (f) with $\beta_N\simeq2$ will be compared in Figs. \ref{figurecaseEF_q}-\ref{figurecaseEF_poincare}.}
  \label{figureHbetaN}
\end{figure}

This paper further investigates the dependence of plasma states on plasma beta ($\beta_N$ and $\beta_p$, the latter is evaluated at half the minor radius). The $q$ profile of the initial equilibria and the current source term remain the same as the base case. However, the pressure and heating sources are scaled down synchronously to obtain equilibria with lower $\beta$. Long-term 3D simulations at different $\beta$ and Hartmann numbers (the magnetic Prandtl number is fixed at 1400) yield different plasma states as shown by Fig. \ref{figureHbetaN}, including flux pumping, sawteeth, and the transition between flux pumping and sawteeth, etc.

\begin{figure}[h]
  \centering
    \includegraphics[width=0.45\textwidth]{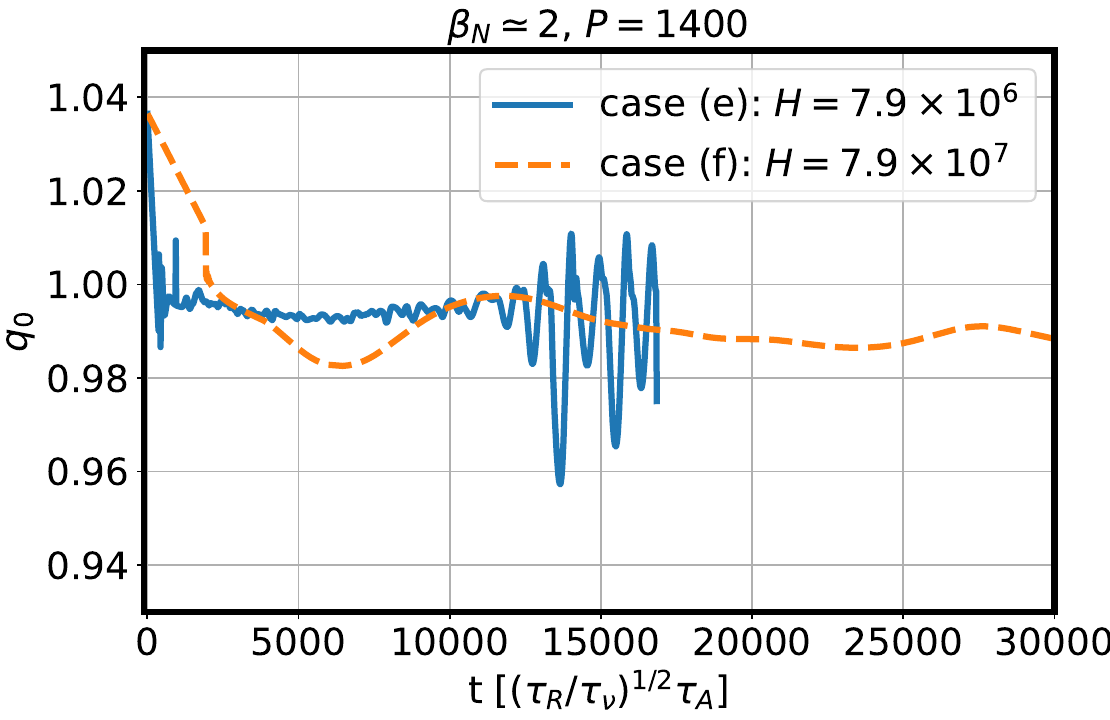}
    \caption{Temporal evolutions of $q_0$ in case (e) and case (f).}
  \label{figurecaseEF_q}
% 2025_0715_scan_visco_eta_Hartmann_Prandtl_correct_heating_rerun_pitagora_scaldown_beta_0.667/Base_case
% plot_multiple_files.py ../00_etax10_viscox10_Hx0.1_Px1/postproc/q_vs_time_at_rhop_eq_0.07.dat postproc/q_vs_time_at_rhop_eq_0.07.dat -legends 'case (e): $H=7.9\times10^6$' 'case (f): $H=7.9\times10^7$' -xm 57.8368999421631 -xl 't [$(\tau_R/\tau_\nu)^{1/2}\tau_A$]' -yl '$q_0$' -ti '$\beta_N\simeq2$, $P = 1400$' -ylim 0.93 1.05 -xlim -100 30000
\end{figure}

Flux pumping only exists at the upper-right corner of the $\beta$-H diagram, indicating the strict conditions for avoiding sawteeth by self-regulation of plasma (the $\beta$ threshold and low enough system dissipation). Nevertheless, at the experimental dissipation ($H=7.9\times10^7$), flux pumping is retained even at a lower $\beta_N\simeq2$ ($\beta_p\simeq2.2$) in the modelling [case (f), see the dashed line in Fig. \ref{figurecaseEF_q}], which differs from the experimental operating window (sawteeth were observed when $\beta_N\simeq2$ in the AUG discharge \#36663) \cite{Burckhart2023NF}. Further lowering $\beta$ leads to the transition from flux pumping to small sawteeth at $\beta_N\simeq 1$ ($\beta_p\simeq1.1$), as indicated by the yellow rectangle. In this state, the core plasma exhibits periodic oscillations resembling small sawteeth, and $q_0$ is oscillating between 1.0 and slightly below 0.98 (not shown). Compared with the AUG experiments, the lower threshold of $\beta$ at $H=7.9\times10^7$ for accessing flux pumping in current JOREK modelling is attributed to the single-fluid physical basis \revisioncolor{that was} used in these studies to limit computational costs. In contrast, the ongoing two-fluid modelling \revisioncolor{\cite{Halpern2011PoPSawtooth, antlitz2025comparisonmhdgyrokineticsimulations, King2011PoPRFP, Ding2004PRLMST_Halldynamo}} and hybrid modelling (with kinetic EPs) \cite{Kolesnichenko2007popEP_QSI, Bogaarts2022JOREK_Kinetic, Zhang2025PPCF_EP_kink, Zhang2026NF_ST_EP} would be essential for precisely demonstrating the threshold of flux pumping, though these become increasingly challenging at low dissipations. 

\begin{figure}[htbp]
  \centering
    \begin{tikzpicture}
    \node[inner sep=0pt] (figs) at (0,0){
    \includegraphics[width=0.45\textwidth]{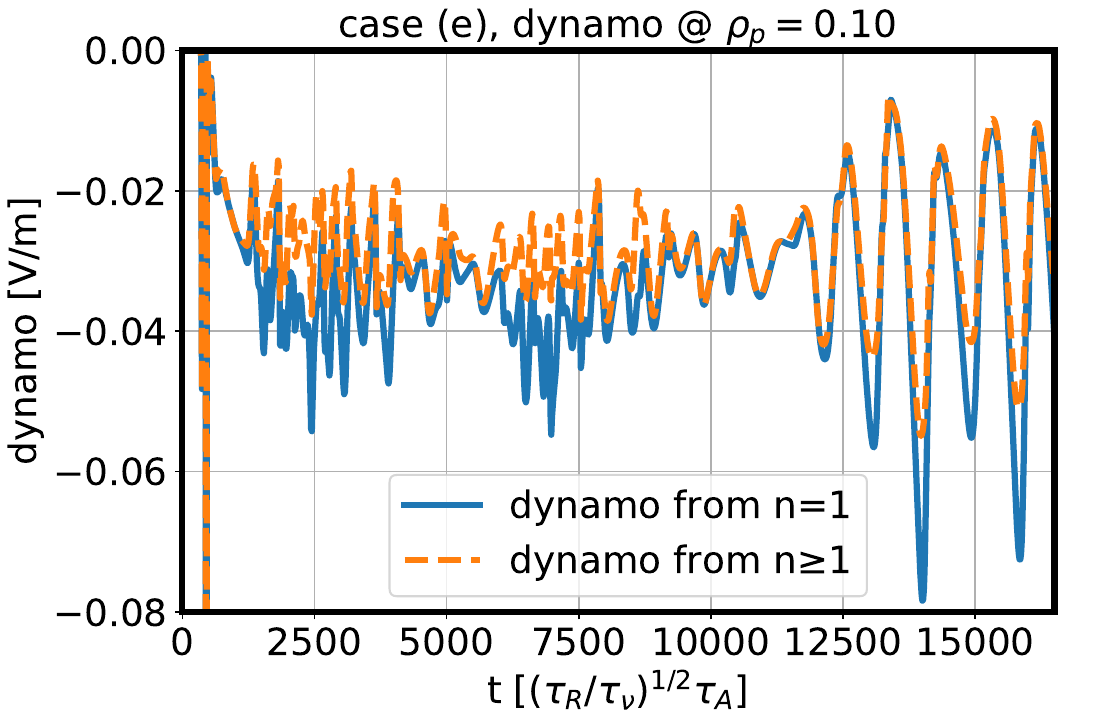}
    \includegraphics[width=0.45\textwidth]{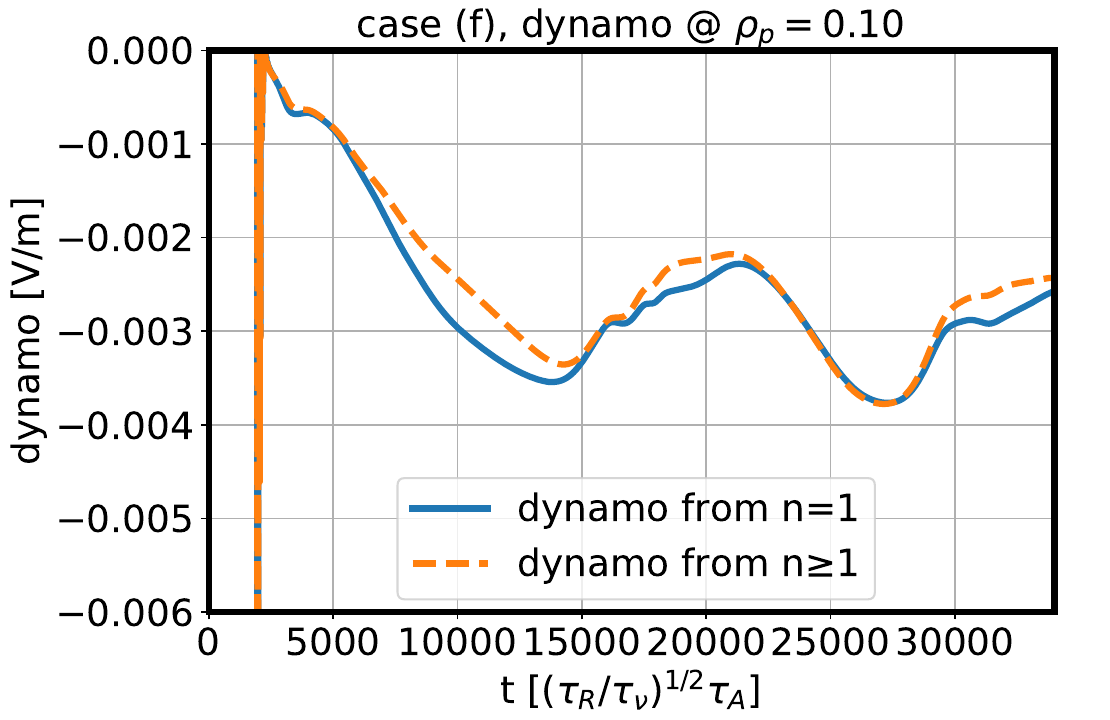}
    };
    \node[align=center,fill=none] at (-4.5, 1.5) {\textbf{(a)}};
    \node[align=center,fill=none] at ( 3.5, 1.5) {\textbf{(b)}};
    \end{tikzpicture}
    \caption{Temporal evolutions of the dynamo electric field generated by the n = 1 component (solid line) and all n $\ge$ 1 components (dashed line) for case (e) (left) and case (f) (right), respectively. }
  \label{figurecaseEF_dynamo_E}
% [00_etax10_viscox10_Hx0.1_Px1]# plot_multiple_files.py postproc/VXB_para_vs_time_at_rho_p_eq_0.10.dat postproc/VXB_para_all_vs_time_at_rho_p_eq_0.10.dat -le 'n=1' 'all' -xm 57.8368999421631 -xl 't [$(\tau_R/\tau_\nu)^{1/2}\tau_A$]' -yl 'dynamo [V/m]' -ti 'case (e), dynamo @ $\rho_p = 0.10$' -le 'dynamo from n=1' 'dynamo from n≥1' -ylim -0.08 0 -xlim 0 16500
% [Base_case]# plot_multiple_files.py postproc/VXB_para_vs_time_at_rho_p_eq_0.10.dat postproc/VXB_para_all_vs_time_at_rho_p_eq_0.10.dat -le 'n=1' 'all' -xm 57.8368999421631 -xl 't [$(\tau_R/\tau_\nu)^{1/2}\tau_A$]' -yl 'dynamo [V/m]' -ti 'case (f), dynamo @ $\rho_p = 0.10$' -le 'dynamo from n=1' 'dynamo from n≥1' -ylim -0.006 0 -xlim 0 34000
\end{figure}

Cases (a) and (e) in Fig. \ref{figureHbetaN} have been further simulated for more than $1.5\times10^4$ $(\tau_R/\tau_\nu)^{1/2}\tau_A$ ($\sim 260$ ms), which is much longer than the typical timescale for identifying flux pumping (showing substantial deviation from the 2D \revisioncolor{simulation}) at the given parameters [$<5000$ $(\tau_R/\tau_\nu)^{1/2}\tau_A$]. In these two cases, the core plasma first evolves into flux pumping and then transitions into small sawtoothing state. To understand the different plasma dynamics in flux pumping and sawtooth, we choose two representative cases (e) and (f) with $\beta_N\simeq2$ in Fig. \ref{figureHbetaN} for comparison ($H=7.9\times10^6$ vs. $7.9\times10^7$). Note that the amplitudes of dynamo and plasma velocities (below) are restored to SI units in this subsection, since the parameters of cases (e) and (f) are close to or at the experimental range. 

The temporal evolutions of $q_0$ for both cases are presented in Fig. \ref{figurecaseEF_q}. For case (e) with higher dissipation, the flux pumping stage is accompanied [before $10^4\ (\tau_R/\tau_\nu)^{1/2}\tau_A$] by weak and high frequency oscillations in $q_0$. After $10^4\ (\tau_R/\tau_\nu)^{1/2}\tau_A$, the oscillation amplitude increases suddenly, where the minimum $q_0$ decreases substantially below unity. In comparison, $q_0$ of the flux pumping case (f) evolves more slowly and saturates around 0.99. Similar evolution characteristics are observed in the temporal evolution of the central dynamo electric field, as shown in Fig. \ref{figurecaseEF_dynamo_E} (a). For case (e), there is a significant difference between the dynamo electric fields respectively generated by the n = 1 component and all n $\ge$ 1 components. Especially during the flux pumping stage of case (e), the dynamo amplitude is significantly reduced by the n $\ge$ 2 components (by up to 50\%). Nevertheless, after the breakdown of flux pumping, the dynamo is again dominated by the n = 1 component, because the complete m/n = 1/1 magnetic reconnection dominates the Kadomtsev-like sawtooth oscillation stage ($q_0$ recovers to above unity after each collapse). In contrast, the dynamo electric field from the n = 1 component is always dominant in the whole simulation of case (f), which results in the two curves almost overlapping in Fig. \ref{figurecaseEF_dynamo_E} (b).

\begin{figure}[htbp]
  \centering
    \begin{tikzpicture}
    \node[inner sep=0pt] (figs) at (0,0){
    \includegraphics[width=0.27\textwidth]{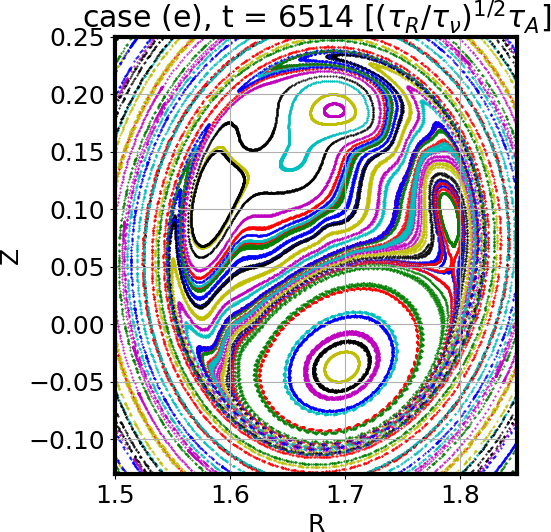}
    \includegraphics[width=0.34\textwidth]{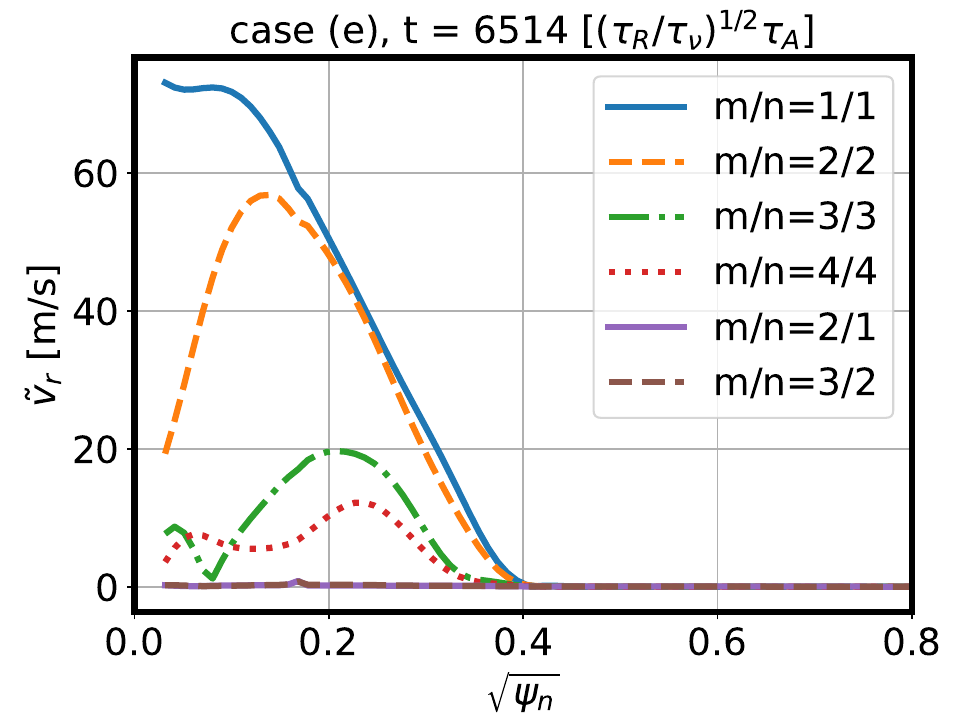}
    \includegraphics[width=0.34\textwidth]{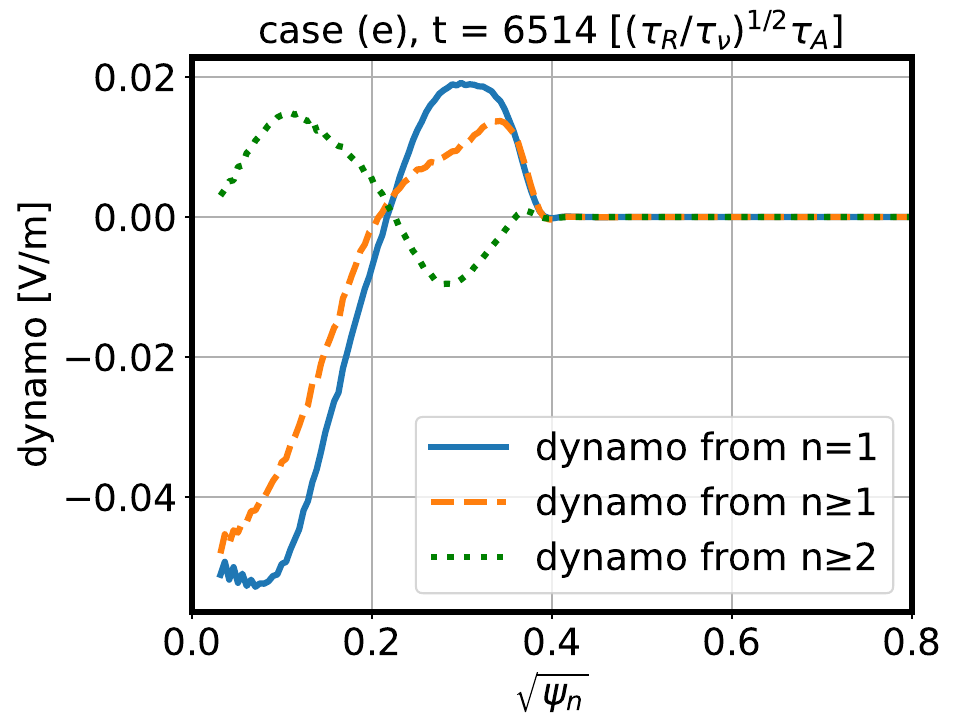}
    };
    \node[inner sep=0pt] (figs) at (0,-4){
    \includegraphics[width=0.27\textwidth]{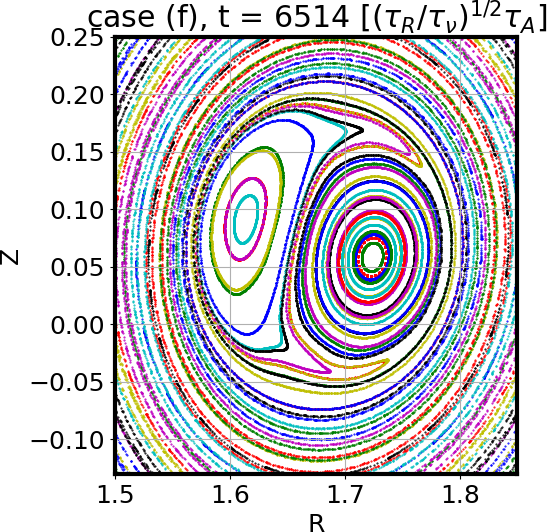}
    \includegraphics[width=0.34\textwidth]{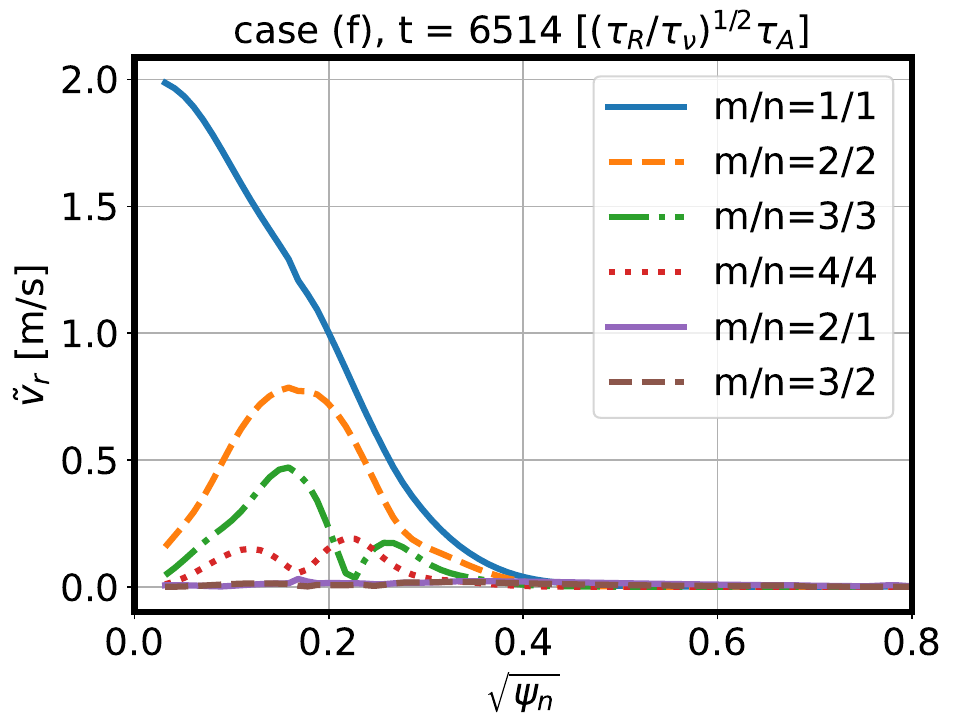}
    \includegraphics[width=0.34\textwidth]{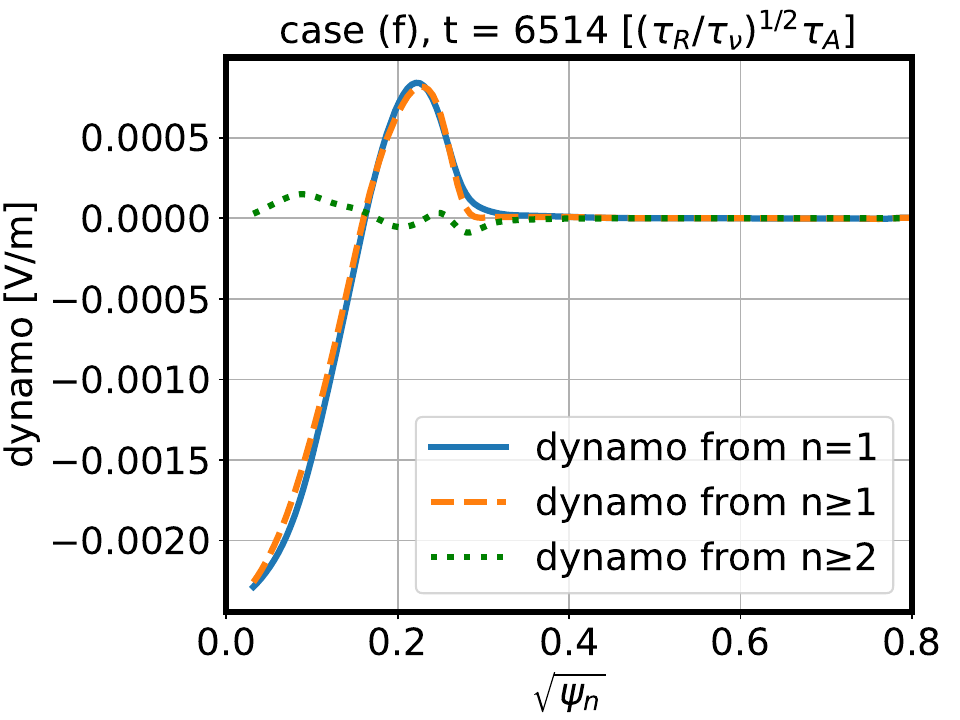}
    };
    \node[align=center,fill=white] at (-6.35,1.35) {\textbf{(a)}};
    \node[align=center,fill=none] at (-1.3,1.2) {\textbf{(b)}};
    \node[align=center,fill=none] at ( 5.8,1.2) {\textbf{(c)}};
    \node[align=center,fill=white] at (-6.35,-2.65) {\textbf{(d)}};
    \node[align=center,fill=none] at (-1.3,-2.8) {\textbf{(e)}};
    \node[align=center,fill=none] at ( 5.8,-2.8) {\textbf{(f)}};
    \end{tikzpicture}
    \caption{For case (e) (upper) and case (f) (lower): Poincaré plots (left), radial mode structures of $\Tilde{v}_r$ with different harmonics (middle), and the dynamo electric field (right) generated by the n = 1 component (solid line), by all n $\ge$ 1 components (dashed line), and \revisioncolor{by} all n $\ge$ 2 components (dotted line) \revisioncolor{are plotted}, respectively.}
  \label{figurecaseEF_poincare}
\end{figure}

The fragility of flux pumping and its final breakdown in case (e) are mainly caused by the emergence of n $\ge$ 2 MHD instabilities. The Poincaré plots, the radial mode structures of $\tilde{v}_r$ with difference m/n components, and the radial profiles of the dynamo electric field are presented in Fig. \ref{figurecaseEF_poincare} at the same time for both cases. As shown by Fig. \ref{figurecaseEF_poincare} (a), for case (e), the plasma core consists of a dominant m/n = 1/1 magnetic island and multiple smaller flux tubes, which alter the central magnetic topology over time. The radial mode structures in Fig. \ref{figurecaseEF_poincare} (b) reveal that the amplitude of the m/n = 2/2 component is comparable to that of the m/n = 1/1. Meanwhile, the n $\ge$ 1 components (dominated by m/n = 2/2) significantly modulate the final dynamo profile (mostly by reducing it) as in Fig. \ref{figurecaseEF_poincare} (c). Nevertheless, in case (f), the mode structure is always dominated by the quasi-stationary m/n = 1/1 component, as shown by Fig. \ref{figurecaseEF_poincare} (d-e). The radial profiles of the dynamo electric field generated by the n = 1 component and all n $\ge$ 1 components are almost identical in Fig. \ref{figurecaseEF_poincare} (f).

The comparisons above show that the flux pumping achieved at the lower dissipation with a dominant m/n = 1/1 convection cell is much more robust than that at higher dissipation, where the latter consists of a more complex mode spectrum. The higher harmonics tend to reduce the amplitude of the total dynamo electric field and also modulate its profile. In this case, to achieve flux pumping, the negative dynamo from the m/n = 1/1 mode has to balance both positive contributions of the external current drive or auxiliary heating and of the high-n harmonics. The residue of the cancellation results in the high-frequency oscillations of $q_0$ and dynamo electric field, which easily leads to the failure of flux pumping and the subsequent sawtooth onset.

\section{\revisioncolor{Relationship with experimental parameters}}\label{SecExpParameter}
\revisioncolor{In this section, the relationship between the above flux pumping regions (expressed by Hartmann number, kinematic viscosity, and resistivity) and the experimental plasma parameters in terms of electron temperature ($T_e$ in eV) and electron density ($N_e$) is discussed. In \cite{Vivenzi2022Viscosity}, different models of viscosity are introduced, including the classical Braginskii viscosities which consider binary Coulomb collisions (parallel $\nu_\parallel$, perpendicular $\nu_\perp$, gyro-viscous $\nu_\times$) \cite{braginskii1965transport, richardson20192019}, the ion temperature gradient (ITG) viscosity $\nu_\text{ITG}$ generated by the damping effect from ion temperature gradient driven mode \cite{Guo1994ViscoITGPOP}, and the Finn viscosity $\nu_\text{Finn}$ due to sound wave propagation along stochastic magnetic field \cite{Finn1992Viscosity}.}

\revisioncolor{With the given AUG parameters, the typical values of these viscosities are $\nu_\parallel\simeq4.2\times10^9\text{m}^2/\text{s}$, $\nu_\perp\simeq3.2\times10^{-3}\text{m}^2/\text{s}$, $\nu_\times\simeq2.2\times10^3\text{m}^2/\text{s}$, $\nu_\text{ITG}\simeq4.9\text{m}^2/\text{s}$, and $\nu_\text{Finn}\simeq2.3\text{m}^2/\text{s}$. The classical Braginskii viscosities are either much higher ($\nu_\parallel$ and $\nu_\times$) or much lower ($\nu_\perp$) than the adopted viscosity in JOREK ($2.7\text{m}^2/\text{s}$), as well as than the ITG and Finn viscosities. Specifically, the classical perpendicular viscosity $\nu_\perp$ is negligible compared with the ITG and Finn viscosities. On the other hand, the classical parallel viscosity $\nu_\parallel$ mainly influences the variation of parallel velocity along the magnetic field lines, which is not the main contributing factor in the dynamo effect under consideration. The gyro viscosity $\nu_\times$ is also much larger than the typical viscosity values for flux pumping (of the order of $1\text{m}^2/\text{s}$), which would normally indicate sawtooth onset based on the above parameter scans. However, the gyro-viscous stress force is essentially non-dissipative because it is perpendicular to the fluid velocity \cite{braginskii1965transport}. Furthermore, in the macroscopic momentum balance, the gyro-viscous force is largely cancelled by the diamagnetic advection term through the well-known \textit{gyro-viscous cancellation} if the ion diamagnetic drift is retained \cite{Schnack2006PoP, Strumberger2023JPP}, which implies that the net effect of these finite Larmor radius (FLR) terms is much smaller than the individual magnitude of $\nu_\times$. Based on the above parameter studies, the ITG and Finn viscosities best match the viscosity range of flux pumping and will therefore be used as the references to estimate the density and temperature ranges for flux pumping.}

\revisioncolor{According to \cite{Guo1994ViscoITGPOP, Vivenzi2022Viscosity}, the ITG kinematic viscosity can be expressed by 
\begin{equation}\label{eq3muITG}
  \begin{split}
    \nu_\text{ITG}=1.08\times10^{-4}\dfrac{\gamma^{1/2}T_eT_i^{1/2}}{Z_\text{eff}L_{T}^{3/4}L_{B}^{1/4}B_0^2},
  \end{split}
\end{equation}
where $e$ is the elementary charge, $T_i$ is the ion temperature in eV, $Z_\text{eff}\simeq1$ is the effective ion charge, $L_{T}\sim1\text{m}$ is the characteristic scale length of temperature, $L_{B}\sim R\simeq1.7\text{m}$ is the characteristic scale length of magnetic field, and $\gamma$ = 1 is the mass number for hydrogen discharges. The Finn kinematic viscosity \cite{Finn1992Viscosity, Vivenzi2022Viscosity} is approximated by
\begin{equation}\label{eq4muFinn}
  \begin{split}
    \nu_\text{Finn}=\sqrt{\dfrac{\gamma_eZ_\text{eff}eT_e+\gamma_ieT_i}{m_i}}L_C\sum_{m,n}\left(\dfrac{b_{m,n}^r}{B_0}\right)^2,
  \end{split}
\end{equation}
where $\gamma_e=1$, $\gamma_i=3$, $m_i$ is the ion mass, $L_C\sim qR\simeq1.7\text{m}$ is the correlation length, and $b_{m,n}^r$ is the magnitude of magnetic field perturbation for a given mode number ($m,n$). From JOREK modelling, $b_{1/1}^r/B_0\sim10^{-3}$ during the flux pumping state. The Spitzer resistivity \cite{Vivenzi2022Viscosity} is given by 
\begin{equation}\label{eq5etaSp}
  \begin{split}
    \eta_\text{Spitzer}=0.51\dfrac{m_e}{N_ee^2\tau_{ei}}=0.51\dfrac{e^2m_e^{1/2}Z_\text{eff}\ln{\Lambda}}{6\sqrt{2}\pi^{3/2}\varepsilon_0^2(eT_e)^{3/2}},
  \end{split}
\end{equation}
where $\tau_{ei}$ is the electron-ion collision time, the Coulomb logarithm $\ln{\Lambda}=\ln{[4\pi(\varepsilon_0 eT_e)^{3/2}/(e^3N_e^{1/2})]}$, $m_e$ is the electron mass, and $\varepsilon_0$ is the vacuum permittivity. }

\revisioncolor{By assuming $T_e\simeq T_i$ and $Z_\text{eff}\simeq 1$, the Hartmann number based on ITG viscosity and Spitzer resistivity [$H_\text{ITG}=\sqrt{\mu_0/\left(\eta_\text{Spitzer}\nu_\text{ITG}\right)}Lv_A$] scales as 
\begin{equation}\label{eq6HITG}
  \begin{split}
    H_\text{ITG}\simeq\sqrt{\dfrac{3(4\pi\varepsilon_0)^2L^2L_T^{3/4}L_B^{1/4}B_0^4}{0.51\times1.08\times10^{-4}\times4\sqrt{2\pi}\gamma^{1/2}e^{1/2}m_e^{1/2}m_iN_e\ln{\Lambda}}}\propto\sqrt{\dfrac{1}{N_e\ln{\Lambda}}}.
  \end{split}
\end{equation}
The Hartmann number based on Finn viscosity and Spitzer resistivity [$H_\text{Finn}=\sqrt{\mu_0/\left(\eta_\text{Spitzer}\nu_\text{Finn}\right)}Lv_A$] scales as 
\begin{equation}\label{eq7HFinn}
  \begin{split}
    H_\text{Finn}\simeq\sqrt{\dfrac{3(4\pi\varepsilon_0)^2L^2B_0^4T_e}{0.51\times8\sqrt{2\pi}\times L_C\sum_{m,n}(b_{m,n}^r)^2em_e^{1/2}m_i^{1/2}N_e\ln{\Lambda}}}\propto\sqrt{\dfrac{T_e}{N_e\ln{\Lambda}}}.
  \end{split}
\end{equation}
The total Hartmann number $H_\text{ITG+Finn}$ contributed by the total viscosity from $\nu_\text{ITG}$ and $\nu_\text{Finn}$ can be calculated in a similar way, i.e., $H_\text{ITG+Finn}=\sqrt{\mu_0/\left[\eta_\text{Spitzer}\left(\nu_\text{ITG}+\nu_\text{Finn}\right)\right]}Lv_A$. Therefore, the potential operating window of flux pumping can be estimated with $H_\text{ITG} > H_c$, $H_\text{Finn} > H_c$, $H_\text{ITG+Finn}>H_c$ and $\beta_p=2\mu_0\langle p\rangle/\langle B_\theta^2\rangle\simeq8\pi^2(\rho_pa)^2\langle N_eeT_e\rangle/\mu_0I_p^2\propto N_eT_e>\beta_{p,c}$, where $a$ the minor radius, the plasma current $I_p$ is assumed to be fixed, $H_c$ and $\beta_{p,c}$ are the thresholds of Hartmann number and poloidal plasma beta ($\rho_p=0.5$) for accessing flux pumping, respectively.}

\begin{figure}[htbp]
  \centering
    \includegraphics[width=0.55\textwidth]{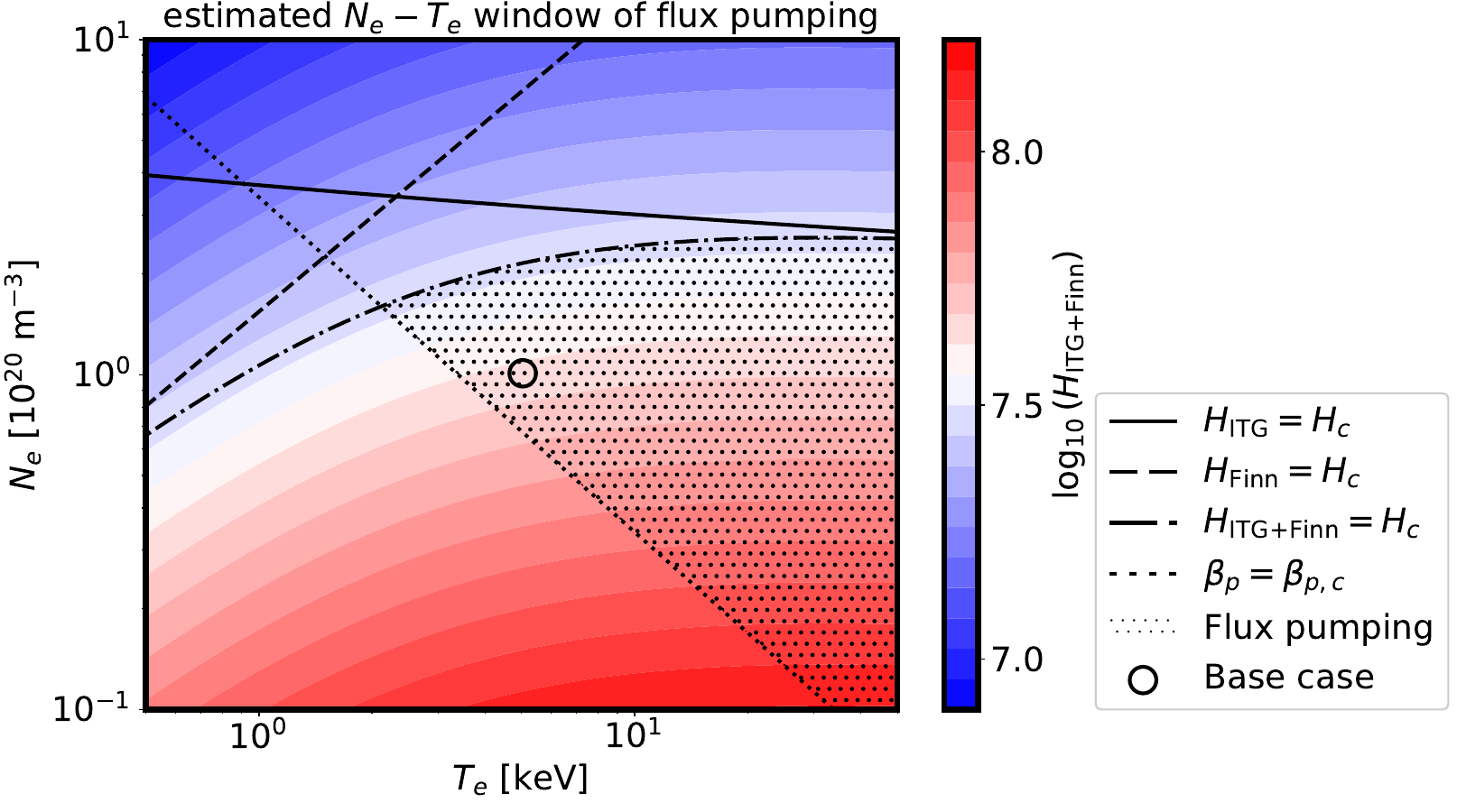}
    \caption{\revisioncolor{The estimated operating window in $N_e-T_e$ for accessing flux pumping. The solid line shows the boundary of $H_\text{ITG} = H_c$. The dashed line shows the boundary of $H_\text{Finn} = H_c$. The dash-dotted line shows the boundary of $H_\text{ITG+Finn} = H_c$. The dotted line shows the boundary of $\beta_p=\beta_{p,c}$. The shaded region indicates the estimated operating window for accessing flux pumping. The black circle shows the parameters of the base case. The color of the contour plot represents the logarithm of $H_\text{ITG+Finn}$. Note that the critical Hartmann number and poloidal plasma beta ($\rho_p=0.5$) are chosen by $H_c\simeq 3\times10^7$ and $\beta_{p,c}\simeq 2.0$, respectively.}}
  \label{figureNeTeWindow}
\end{figure}

\revisioncolor{From the above numerical parameter studies, the critical Hartmann number is chosen by $H_c\simeq 3\times10^7$, and the threshold poloidal plasma beta ($\rho_p=0.5$) is $\beta_{p,c}\simeq 2.0$. The estimated operating window of flux pumping in $N_e-T_e$ space is plotted by the shaded region in Fig. \ref{figureNeTeWindow}. In general, flux pumping mainly exists in the high temperature regime, which corresponds to a low Spitzer resistivity. However, the Hartmann number, determined by Finn and ITG viscosity models, establishes a critical density upper bound. Beyond this limit, the decrease in $H$ increases the relative magnitude of the dissipative terms as shown in Eqs. \ref{eq1induction} and \ref{eq2momentum}. This results in the increased viscous damping of convective velocity and an enhanced magnetic diffusion, thereby pushing the system below the threshold for efficient flux pumping. On the other hand, the threshold of plasma beta provides the lower limit of density, below which the plasma pressure gradient may not be strong enough to drive the quasi-interchange mode and sustain the flux pumping state. Nevertheless, the considered parameter ranges of $N_e$ and $T_e$ are much larger than the typical operating window of AUG discharges. This is because the proposed model based on the Hartmann number can only provide a qualitative estimate of the rough parameter region of flux pumping, as shown by Fig. \ref{figureNeTeWindow}. In practice, the realistic effective viscosity values and the nonlinear dynamics of MHD instability depend on more parameters than just $N_e$ and $T_e$. Therefore, nonlinear 3D simulations or a well-calibrated surrogate model being developed \cite{Krebs2023EPSreduced} remain necessary for precisely predicting the feasibility of flux pumping.}

% }

\section{Conclusion and discussion}\label{secSummary}
In this work, the flux pumping discharge (\#36663) of the AUG hybrid scenario has been systematically modelled based on the two-temperature, full MHD model of JOREK. The study commences with a comparison of 2D/3D nonlinear simulations at realistic parameters with the experimental observations. The simulation reproduces the radial redistribution of current density in the plasma center due to the continuous dynamo electric field generated by the m/n = 1/1 MHD mode. The dynamo of the order of mV/m is negative in the core and positive at $\rho_p\gsim0.2$, which clamps the flat current density at the amplitude of 2.5 $\text{MA/m}^2$ and the central shear-free $q$ profile around 1.0, thereby preventing sawtooth onset. The simulation results at AUG parameters are quantitatively consistent with the experimental observations in terms of current redistribution and dynamo electric field in the presence of flux pumping.

In this context, parameter scans have been performed to investigate plasma bifurcations at different system dissipations and plasma betas. By increasing the resistivity and viscosity, equivalent to decreasing the Hartmann number, four regimes of core plasma states are identified. First, flux pumping is obtained at extremely low dissipation ($H\gsim 10^6$, well beyond the typical range modelled for tokamak or RFP plasmas, i.e., $H\lsim10^5$). Nevertheless, flux pumping is found to be much less robust due to the destabilization of n $\ge$ 2 components at the lower Hartmann number ($\sim10^6$), which eventually leads to sawtooth onset in the long-term 3D simulations. Secondly, sawtooth oscillations are observed at moderate dissipations ($10^5\lsim H\lsim10^7$). The intermittent dynamo electric field is responsible for raising $q_0$ to unity during the core relaxation collapse. Thirdly, at the higher dissipation ($10^4\lsim H\lsim10^6$), \revisioncolor{a} single giant sawtooth crash state followed by a quasi-stationary m/n = 1/1 resistive internal kink mode is observed. The later stage is similar to flux pumping with a continuous dynamo offsetting the current drive, during which $q_0$ ($\sim 0.9$) is relatively lower than unity but much higher than that of the 2D case. The final state of the highest dissipation ($H\lsim10^4$) corresponds to the complete failure of the dynamo to sustain or modulate the $q$ profile (the dynamo amplitude is at noise level). Therefore, $q_0$ directly decreases to the value ($\sim0.7$) close to the 2D case. 

The bifurcated plasma behaviours at different Hartmann numbers are qualitatively consistent with existing MHD modelling of FRP and tokamak plasmas. This work bridges the previous MHD modelling of sawtooth or similar plasma relaxation events at moderate Hartmann number \cite{Krebs2017PoPFluxPumping, Shen2018NF, Zhang2020NFfluxpumping, Cappello2000PRLRFP, Cappello2006PoPRFP, Bonfiglio2013PRL, Veranda2019NFRFX, Piovesan2017NF} and the latest quantitative simulation of flux pumping at experimentally relevant high Hartmann numbers \cite{Zhang2025NFfluxpumping}. Meanwhile, the plasma beta threshold for accessing flux pumping in full MHD modelling is evaluated. The result is qualitatively consistent with the AUG experiments, but also suggests the necessity of improving the extended simulation model, in particular by including two-fluid effects \revisioncolor{\cite{Halpern2011PoPSawtooth, antlitz2025comparisonmhdgyrokineticsimulations, King2011PoPRFP, Ding2004PRLMST_Halldynamo}}, the kinetic EPs \cite{Kolesnichenko2007popEP_QSI, Bogaarts2022JOREK_Kinetic, Zhang2025PPCF_EP_kink, Zhang2026NF_ST_EP}, and the relevant fishbones \cite{Burckhart2023NF, Guenter1999FB}. \revisioncolor{The two-fluid based extended MHD model could introduce finite stabilization effects on the dominant mode, and allow the self-consistent inclusion and analysis of the Hall and diamagnetic dynamo effects if the generalized Ohm's law is retained \cite{King2011PoPRFP, Ding2004PRLMST_Halldynamo, JI2001alphaDynamoRFP, Mao2023PRR}, thereby influencing the parameter regimes of flux pumping.} \revisioncolor{In the end, based on the anomalous ITG and Finn viscosity models \cite{Guo1994ViscoITGPOP, Finn1992Viscosity, Vivenzi2022Viscosity}, as well as the critical Hartmann number and plasma beta for accessing flux pumping, the rough operating window of AUG discharges in terms of plasma density and temperature is estimated. Flux pumping is found likely to exist at regimes of higher temperature and moderate density. Future studies will involve more detailed direct comparisons between the numerical modelling results and the experimental database of discharges.}

In parallel, the studies involving the scan of heat and current sources, the EP/fishbone effects, the modelling targeting JET flux pumping experiments \cite{Burckhart2024EPS, Alex2025IAEA}, and the extended MHD development are being conducted. These efforts are critical for understanding flux pumping through direct 3D MHD simulations and for calibrating the fast surrogate model being developed \cite{Krebs2023EPSreduced}. The surrogate model aims to efficiently predict the dynamo amplitude and assess the feasibility of flux pumping in existing tokamaks like AUG and JET, and future larger devices like ITER and DEMO.

\section*{\textbf{Acknowledgements}}
The author H. Zhang would like to acknowledge J. Stober for reviewing the original manuscript for IAEA FEC 2025 and for providing constructive suggestions. 
The simulations were finished mainly on the supercomputers (Marconi-Fusion, Leonardo, Pitagora) hosted by CINECA in Italy, the JFRS-1 supercomputer at IFERC-CSC in Japan, and the TOK cluster at the Max Planck Institute for Plasma Physics in Germany. This work has been carried out within the framework of the EUROfusion Consortium, funded by the European Union via the Euratom Research and Training Programme (Grant Agreement No 101052200 EUROfusion). Views and opinions expressed are however those of the author(s) only and do not necessarily reflect those of the European Union or the European Commission. Neither the European Union nor the European Commission can be held responsible for them.

\section*{\textbf{References}}\vspace{-1.2em}
\bibliography{References}

\end{document}